\documentclass[12pt,twoside]{book}

\usepackage{rscsoftmat}
\usepackage{amssymb}
\usepackage{graphicx}
\usepackage{booktabs}
\usepackage{bm}
\usepackage{mathtools}
\usepackage{mathrsfs}
\usepackage{amsfonts}
\usepackage{amsthm}
\usepackage{xcolor}
\usepackage{soul}

\usepackage{latexsym}
\usepackage{multirow}

\newcommand{\chapterauthor}{ L. N. Carenza $^{a}$, G. Gonnella$^{b,\star}$, G. Negro$^{b}$} 
\newcommand{\authoraffiliation}{$^a$ Instituut-Lorentz, Universiteit Leiden, P.O. Box 9506, 2300 RA Leiden, Netherlands, \\
$^b$Dipartimento di Fisica, Università degli Studi di Bari and INFN, Sezione di Bari, via Amendola 173, Bari, I-70126, Italy} 
\newcommand{\email}{giuseppe.gonnella@ba.infn.it} 

\newcommand{\abstract}{
 In the last years self-motile droplets attracted the attention of scientists from different fields ranging from applied biology to theoretical physics, because of their promising technological applications and important biological implications. 
In this Chapter we review the state of the art of the research on active  droplets with a particular focus on theoretical and numerical studies. In particular, we reviewed the active gel theory, namely a generalization of the standard Landau-de Gennes theory for liquid crystals adapted to take into account internal active injection due to the presence of self-motile constituents. 
When confined in finite geometries, liquid crystalline-like systems are also subject to topological constraints. Because of the relevance of topology in many different realizations of active droplet\textcolor{black}{s}, we also reviewed some fundamental topological concepts. We review how motility arises in different realizations of active droplet both in $2d$ and $3d$ as the result of the breaking of specific symmetries, by looking in particular detail \textcolor{black}{at} the case of polar and nematic droplets and shells of active liquid crystal.
}

\begin{document}

\chapter{Motility and self propulsion  of active droplets}

\MiniToC
\section{Introduction}
\label{sec:1}
Droplets, vesicles and cells with motile properties have recently attracted much attention in the scientific community spanning from biophysics to condensed matter and chemical physics. The interest in this particular kind of realizations of active matter concerns the relevance which self-motile droplets have in many biological processes at the base of life and the possibility to replicate such processes artificially. Let us mention a few examples. For instance, vesicles are implemented by cells as \emph{cargo} to transfer substances from a source to a target organelle (compartments in a cell in which specific activities --\emph{e.g.} synthesis of a certain enz\textcolor{black}{y}me-- takes place). Moreover, extracellular vesicles likely play a fundamental role in coagulation and intercellular signalling~\cite{vandelpol2012}. Understanding this complex machinery of intra- and extra-cellular messaging is of drastic importance in medicine and for the fight against many diseases, as witnessed by the Nobel Prize in Medicine awarded in 2012 to James Rothman, Randy Schekman and Thomas S\"{u}dhof for their studies on this topic.
Cells themselves can be seen as self-propulsive units under many circumstances. For instance, bacteria among prokaryota may be equipped with \emph{flagella}, long filamentous protrusions attached to the cellular membrane which bacteria exploit to move in a fluidic environment.
Recently, synthetic vesicles and shells of bio-inspired material were also developed in the lab and used to perform experiments, paving the way towards the development of synthetic microdevices and micromotors which could constitute an epoch-making change in medical therapies and development of cyborg technology.

Motivated by the biological and medical implications, research on self-motile droplets and vesicles found in the statistical mechanics of non-equilibrium systems a natural background for the development of theoretical models able to capture the gargantua of different behaviors observed in nature. Among many models proposed in the field, one of the most appealing and successful is the \emph{active gel theory} that happens to be particularly useful to describe dense filamentous systems which order in a liquid crystalline fashion and have been widely studied in experiments. 
These realizations of active droplets which develop nematic or polar order also brought to light the relevance of topology in the context of biological processes such as morphogenesis~\cite{Livshits2017,Maroudassacks2021,Hoffmann2021}. 

The purpose of this Chapter will be to twofold. On the one hand we will review a number of topical results on droplet motility with a special emphasis on analytical and simulation studies. On the other hand, we will also provide an introduction to the active gel theory and to some fundamental topological concepts which come at hand when dealing with spherical (or more complex) geometries. It \textcolor{black}{is worth} mentioning \textcolor{black}{since} the beginning that topological defects --namely regions where order is lost-- play a very relevant role in the onset of self-motile states since they are strongly related to distortions in the liquid crystalline pattern which are in turn the primary responsible of flow instabilities.

Section~\ref{sec:2} is devoted to a review of various experimental realizations of active droplets. The focus of this Section is mostly on the possible routes that can be experimentally practicable to achieve droplet self-propulsion in terms of different elementary constituents and driving physical mechanisms, rather than a comprehensive review on results.

Section~\ref{sec:3} is an introduction to the active gel theory, \emph{i.e.} the continuous theoretical model which is commonly used in the context of theoretical and numerical investigation on dense active matter. After discussing the main equilibrium features of a passive liquid crystal we show that activity can be introduced in this model as a source term in the balance equation of linear momentum. In this respect, we report a topical analytical argument proving that activity may lead to self-sustained flows resulting into non-equilibrium steady states when the active gel is confined in a closed geometry. We also discuss how this mechanism is crucial \textcolor{black}{for} the development of droplet motility and we report the results of some emblematic numerical experiments in 2D.

In Section~\ref{sec:4} we present a review of some topological concepts like topological defects in liquid crystalline systems and how these couple to the geometry where they are confined. The information provided serve\textcolor{black}{s} as introductory material for Section~\ref{sec:5} where the case of active droplets and shells in the full 3D space is considered. Here the physical mechanism leading to droplet motility is interpreted in terms of symmetry breaking and we show how rotational and linear motion in a droplet can be coupled by exploiting chirality of the fundamental constituents.

\section{Review of different experimental realizations of self propelled droplets}

\label{sec:2}

\begin{figure}[htp]
\centering
\includegraphics[width=0.85\textwidth]{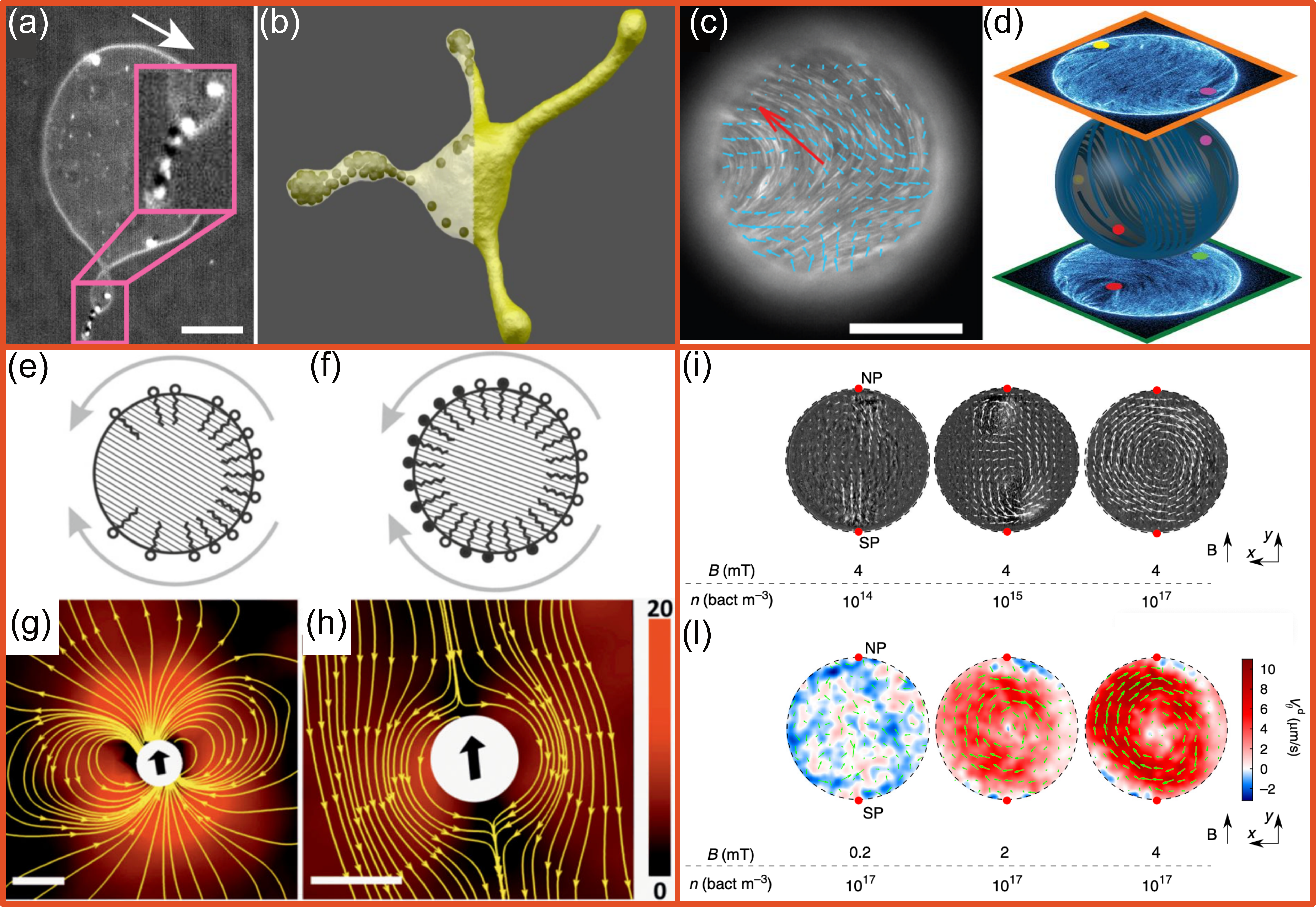}
\caption{\textbf{Different realizations of self-propelled droplets.}  \textbf{(a-b)} Janus colloids encapsulated in vesicles. Bright-field microscopy image (a) and simulations snapshot (b) of  self propelled Janus colloids (\textit{half Pt-coated polystyrene}) encapsulated in giant unilamellar vesicles (GUVs) of \textit{1,2-dioleoyl-sn-glycero-3-phosphocholine lipids}. 
\textbf{(c)-(d)} Confined microtubules bundles (MT).  
\textbf{(c)} Fluorescence image of active MT bundles incapsulated into a water-in-oil droplet. The resulting active liquid crystalline phase  spontaneously adsorbs onto the oil-water interface exhibiting streaming flows, indicated with cyan arrows, eventually leading to droplet translation, signalled by the red arrow. 
\textbf{(d)} Hemisphere projection of a 3D confocal stack of a nematic vesicle. The positions of four $+1/2$ disclination defects are identified by dots. 
\textbf{(e)-(h)} Marangoni self propelled droplets.  
\textbf{(e-f)} Surfactant molecules adsorb at the droplet surface reducing the interfacial tension. Among other panels (e)-(f) show two possibilities to generate Marangoni stresses along the surface of a droplet: \textbf{(e)} inhomogeneous surfactant coverage and \textbf{(f)} a spatially constant surfactant coverage with spatially varying surface activity. Black/white surfactant denote lower/higher surface activity which might be caused by chemical reactions. In both cases, the resulting Marangoni stress points in the direction of increasing surface tension (arrows around the drop).
\textbf{(g-h)} Velocity fields around a droplet as determined by  particle image velocimetry (PIV). Yellow lines denote stream lines of the flow.  The magnitude of the velocity is color coded in microns per second, scale bars denote 100 microns. (f) Velocity field in the laboratory frame and (h) in co-moving droplet frame after subtracting the droplet velocity. 
\textbf{(i)-(l)} Droplets with a suspension of magnetocatics bacteria. 
\textbf{(i)} 
Phase contrast images of water in oil droplets with a suspension of Magnetospirillum gryphiswaldense MSR-1 magnetocatics bacteria  with superimposed with time average velocity fields (withe arrows), for different values of cell density $n$.
\textbf{(l)}
Colored maps of the orthoradial projection of the instantaneous  velocity fields with superimposed instantaneous  velocity field (green arrows) for the same system in (i) for different values of the magnetic field $B$.
\textcolor{black}{Panels (a-b) are adapted from Ref.~\cite{vescicles1} with permission from Springer Nature; panel (c) from Ref.~\cite{sanchez2012} with permission from The American Association for the Advancement of Science; panel~(d) from Ref.~\cite{Keber1135} with permission from The American Association for the Advancement of Science; panels (e-h) from Ref.~\cite{C4SM00550C} with permission from The Royal Society of Chemistry; panels (i-l) from Ref.~\cite{vincenti2019}, CC BY.}
}
\label{fig:1}
\end{figure}

Droplets with self-propulsive ability can be realized in many ways, differing in size, building components and fuelling mechanism. In this Section we will provide an overview on the wide panorama of active droplets, dwelling only on a few topical examples. A first broad classification of self-propelled droplets relies on the physical mechanism which fuels the propulsion. This can arise as the result either of the dynamics of suspended active constituents encapsulated in (or eventually \emph{on}) the droplet, or interfacial effects which generate hydrodynamic stresses and flows able to advect the droplet itself.

\textcolor{black}{In order to achieve droplet motility, the most straightforward idea, relies on the possibity to confine self-propelled particles (SPPs) in a vescicle.} This kind of realization 
has attracted much attention because of the many different dynamical regimes which can be achieved in such a way that may find important applications in designing synthetic cells. For instance, Vutukuri \emph{et al.}~\cite{vescicles1} have recently developed an experimental protocol to control the shape change of a lipid vesicle by confining in its interior self-propelled Janus colloids. These SPPs, whose typical size $\sim 1 \mu$m,  exploit diffusiophoretic effects to propel with a drifting speed which can vary from a few $\mu$m$/s$ up to $\sim 200 \mu$m$/s$ by opportunely setting the fuel concentration --\emph{i.e.} hydrogen peroxide ($H_2O_2$). Interestingly, when active particles land on the surface, they can deform the vesicle giving rise to multifarious dynamical scenarios including a fluctuating regime or the formation of prolate and bola-like shapes, according to the mutual effect of active forcing, concentration of SPP and vesicle surface tension.
In panel~(a) of Fig.~\ref{fig:1} a vesicle is shown, as it develops a tether consisting of multiple active particles, illustrated in the magnified region. The authors of~\cite{vescicles1} also provide \textcolor{black}{insights} into the mechanics of such dendritic growth by means of numerical simulations, showing that tethers formation is first initiated by a few SPPs which slightly deform the surface. As more and more SPPs get trapped in this region of higher curvature the effective active force becomes more and more localized, eventually leading to the growth of the protrusion (see panel~(b)).

\textcolor{black}{Droplet motility was also achieved by making use of cytoskeletal filaments like microtubules (MT). These are found in both prokaryote and eukaryote cells and are rod-like filaments which play a fundamental role in cell division, cellular structure and motility.}
In experiments, they are used in presence of depletion agents like non-absorbing polymers (\emph{i.e.} polyethylene glycol also known as PEG) and arrange in bundles characterized by nematic symmetry with a typical coherence length of $\sim 10 \mu$m. This particular kind of biological matter can be \emph{activated} by dispersing ATP --the fuel of cells-- which is hydrolyzed by the kinesin motors linked to the microtubules to provide the mechanical energy for the nematic bundles to start flowing, a property which was implemented in~\cite{sanchez2012} to give rise to the spontaneous motility of a water-in-oil droplet incapsulating MT bundles, shown in panel (c) of Fig. \ref{fig:1} (with a reference speed of a few $\mu$m$/s$). At large MT concentrations, the motion of the single constituents strongly influences the dynamics of the surrounding bundles whose elastic response gives rise to an effective \emph{active nematics}. 
Keber \emph{et al.}~\cite{Keber1135} were the first ones to study shells of active nematics in the lab by confining such mixture of microtubules, kinesin motors and PEG to the interface of water-in-oil droplets or to the inner leaflet of the lipid bilayer of a vesicle. 
\textcolor{black}{Such preparation is particularly appealing as it represents an example of a physical system whose topological properties can be used to induce morphological changes, paving the way towards the design of mesoscopic metamaterials. }
[We will come back to these and other topological concepts in Sec.~\ref{sec:4}] Interestingly, Keber \emph{et al.} observed \textcolor{black}{that the centre of mass of the droplet did not move significantly from its originary position. However,} the topological defects on the shell (see colored points in panel~(d) of Fig. \ref{fig:1}) oscillated between tetrahedral and planar configurations in a periodic fashion under the action of the active force dipoles fuelling the shearing motion of the MT bundles. Even more surprisingly, by reducing the surface tension of the vesicle, yet another dynamics was observed with thin protrusions of active nematics buckling out of the shell in correspondence of the position of the topological defects. Also, much effort has been spent in trying to disentangle and control the intricate nematodynamics from a theoretical perspective. 
Among the most successful methods, the \emph{active gel theory}, based on a generalisation of the Ericksen theory for liquid crystals to incorporate the non-equilibrium effects of activity, has led to excellent results in recovering many experimental observations and formulate predictions. This approach, also considered in the Chapter of J. Yeomans, will be the \emph{Leitmotiv} of the next Sections.

Another driving mechanisms for self-propelled droplets relies on the interfacial properties of emulsified droplets which can be also exploited to achieve a drifting force. For instance, droplets with nonuniform surface tension may experience the effects of the Marangoni effect in systems with surfactant. 
There are two main possibilities to generate Marangoni stresses along the surface of a droplet, both depicted in Fig.~\ref{fig:1} (e)-(f):  Inhomogeneous surfactant coverage (e), or a spatially constant surfactant coverage with spatially varying surface activity (f). In both cases, the resulting Marangoni stress points in the direction of increasing surface tension (arrows around the drop) so that the resulting droplet motion points into the opposite direction.
The physical origin of such phenomenon can be uncovered by observing that surface tension gradients have the dimensions of a force per unit area, hence of a stress. On the one hand, this gives rise to backflows which advect the droplet itself inducing mass transfer directed in the direction of larger surface tension resulting in the motion of the droplet.
On the other hand, such surface inhomogeneities are not long lived, as the system naturally evolve to restore an uniform configurations. 
It remains clear that the droplet drift can only be initiated by surfactant inhomogeneities which would rapidly smooth out in absence of some other mechanism able to maintain surface tension gradients out of equilibrium. This can be possible exploiting either some kind of chemical reaction able to change the surface activity or by stabilizing surfactant gradients through inhomogeneous depletion in liquid crystalline emulsions.
The typical drifting speed $v \sim R \Delta \gamma / \eta$, where $R$ is the radius of the droplet, $\Delta \gamma$ is the typical variation of the surface tensions and $\eta$ the viscosity of the fluid, as one may easily rationalize through dimensional analysis. 
The typical size of droplets moving due to Marangoni stress is of $\sim 100 \mu$m and their propulsion velocity is of order $\sim 10 \mu$m$/s$, while the typical flow field is depicted in Fig.~\ref{fig:1}(g)-(h) respectively in the reference frame of the droplet and the lab.

Another recent realization of droplets capable of generating self-sustained flows has been obtained experimentally by making use magnetotactic bacteria. These are motile prokaryotes that synthesize chains of lipid-bound, magnetic nanoparticles called magnetosomes, consisting of membrane-enclosed intracellular crystals of a magnetic iron mineral. The chain-like arrangement of magnetosomes confers a magnetic dipole moment to the cell, which enables it to orient and migrate along magnetic field lines \cite{blakemore377}.
\textcolor{black}{Pierce \emph{et al.}~\cite{pierce2017} confined a suspension of Magnetospirillum magneticum AMB-1 in a water-in-oil droplet and showed that the droplet dynamics can be tuned through the use of weak, uniform, external magnetic fields and local, micromagnetic surface patterns.}
\textcolor{black}{Vincenti} \emph{et al.}~\cite{vincenti2019} showed that under the application of a constant magnetic field, motile magnetotactic bacteria confined in water in oil droplets self-assemble into a rotary motor exerting a torque on the external oil phase. A collective motion in the form of a large-scale vortex, reversable by inverting the field direction, builds up in the droplet with a vorticity perpendicular to the magnetic field.  The emergence and the strength of collective vortical motion can be controlled by cell density (see Fig. \ref{fig:1}(i)) and the magnetic field (see Fig. \ref{fig:1}(l)). They also  characterized quantitatively the mechanical energy extractable from this new biological and self-assembled motor, showing how this set-up could serve as the basis for a variety of novel, dynamic, microscale machines operating in low Reynolds numbers environments.

This concludes our brief summary on different experimental realizations of self-propelled droplets. For a more detailed review and discussion on experimental realizations, the interested reader can refer to~\cite{Gompper2020}.


\section{Theoretical models and confined active gels in 2D}
\label{sec:3}
In this Section, we will introduce the active gel theory and how it can be exploited to model active droplets.
For the sake of clarity, we will consider realizations with MT as a target system, \textcolor{black}{for instance as the one shown in Fig.~\ref{fig:1}(c-d)}, and we will explicitly refer to them in the rest of this Chapter, even if the active gel theory has been successfully used to investigate other kinds of active systems  --for instance bacterial or acto-myosin  suspensions-- where the anisotropy of the elementary constituents gives rise to effective aligning interactions.
The key ingredient of this approach is to focus on the description of the mesoscopic scales, relevant for capturing the dynamics of the system in its entirety. This can be done through a coarse-grained approach where only the orientational and elastic features of the system are retained, while the inherent biochemical complexity of the particular realization is traced out, as one only deals with a few hydrodynamic continuous fields, rather than the Lagrangian description of the single constituents~\footnote{\textcolor{black}{For other approaches taking into account the dynamics of the individual constituents we suggest the reader to refer to the following references~\cite{bechinger2016,digregorio2018,cugliandolo2017}.}}.
In the following, we will first introduce the order parameters of the theory and their evolution equations. Afterwards, we will consider a topical result obtained in the scope of the active gel theory uncovering the relation between self-sustained flows and orientational deformations in the pattern of the active constituents. \textcolor{black}{Finally, we will present a  summary on confined active gels in $2D$, to show in this simplified environment how geometrical confinement may lead to droplet motility.}

\subsection{Order Parameters}
A feature common to most aggregates of living constituents is to arrange in an ordered fashion during motion. 
Apart from the most striking examples found in the animal kingdom, many cellular and sub-cellular systems also exhibit important ordering and collective behaviors with relevant topological features which arise from the composite assembling of the constituents into phases which break isotropic symmetry.
To capture the degree of ordering, it is customary to introduce suitable order parameters which respect the symmetries of the system. 
\textcolor{black}{In this Chapter we will only consider systems with polar and nematic symmetry. However, it worths noticing that many biological systems also exhibit other kinds of ordering, \emph{e.g.} hexatic one, whose properties should be captured by an appropriate order parameter~\cite{giomi2021}.}
For instance, an active polar fluid is composed of elongated self-propelled particles, which feature a head and a tail. Such systems may order in polar states, when all the particles are on average aligned along the same direction, as for flagellate bacteria which propel along the direction of their head~\cite{dombrowski2004}. 
The continuum characterization of a dense suspension of anisotropic particles may either emerge from a coarse grained description of a microscopical model~\cite{degennes1993} or from a theory
based on general symmetry arguments~\cite{martin1972,beris1994}. Following the former approach for a system of arrow-like
particles (with broken \textcolor{black}{head}-tail symmetry),  the probability density function $f_P(\bm{\nu}, \mathbf{r},t)$ can be introduced as the probability to find a particle oriented along the vector $\bm{\nu}$ at position $\mathbf{r}$ and time $t$.
This can be normalized by requiring the concentration 
of the suspended particles $\phi(\mathbf{r},t)$ to be given by the following relation:
\begin{equation}
\phi(\mathbf{r},t)= \int \mathrm{d \Omega}  f_P(\bm{\nu}, \mathbf{r},t),
\label{eqn:concentration_field}
\end{equation}
where the integration is performed over the solid angle $\mathrm{\Omega}$.
The polarization field can be defined as 
\begin{equation}
\mathbf{P}(\mathbf{r},t) = \langle \bm{\nu}  \rangle  = \dfrac{\int \mathrm{d \Omega} f_P(\bm{\nu}, \mathbf{r},t) \bm{\nu}}{\int \mathrm{d \Omega}  f_P(\bm{\nu}, \mathbf{r},t)}=  P(\mathbf{r},t) \mathbf{n}(\mathbf{r},t),
\end{equation}
where $\mathbf{n}(\mathbf{r},t)$ is a unit vector defining the local mean orientation of particles in the neighborhood of $\mathbf{r}$, and  $P(\mathbf{r},t)$ is a measure of the local degree of alignment, ranging from $0$ (in an isotropic state) to $1$ (in a perfectly polarized state).

However, systems of intrinsic polar particles, may still restore head-tail symmetry, when interactions favor alignment regardless of the
polarity of the individual particles. For instance, this is the case for MT which have their intrinsic polarity, while the resulting bundle is basically apolar and the final result is a nematic network of filaments with \textcolor{black}{head}-tail symmtetry. the  nematic phase cannot be described by a vector field, as both orientations $\bm{\nu}$ and $-\bm{\nu}$ equally contribute to the same ordered state, and the polar order parameter $P(\mathbf{r},t)$ drops to zero. In the case of a system of rod-like particles, the minimal mathematical object able to capture the orientational features of the system is a symmetric rank-2 tensor --namely the Q-tensor, in the language of liquid crystals-- which can be defined as
\begin{equation}
\label{eqn:particle_nematic_tensor}
\mathbf{Q}(\mathbf{r},t) = \langle \bm{\nu} \bm{\nu} - \dfrac{1}{d} \mathbf{I} \rangle  = \dfrac{\int \mathrm{d \Omega} f_Q(\bm{\nu}, \mathbf{r},t) \left(\bm{\nu} \bm{\nu} -\dfrac{1}{d} \mathbf{I} \right)}{\int \mathrm{d \Omega}  f_Q(\bm{\nu}, \mathbf{r},t)},
\end{equation}
where $d$ is the dimensionality of the system and $f_Q(\bm{\nu}, \mathbf{r},t)$ is the probability density to find a nematic particle oriented along $\bm{\nu}$ at position $\mathbf{r}$ and time $t$, invariant under transformations $\bm{\nu} \rightarrow - \bm{\nu}$.
By definition, the Q-tensor is symmetric and traceless, so that it always admits a diagonal representation:
\begin{equation}
\mathbf{Q}^{diag} =\begin{pmatrix}
\dfrac{2}{3}S & 0 & 0\\
0 & -\dfrac{1}{3}S+\psi & 0 \\
0 & 0 & -\dfrac{1}{3} S -\psi
\end{pmatrix},
\label{eqn:diagQ}
\end{equation}
where $S$ and $\psi$ are  two scalar order parameters, respectively measuring the degree of uniaxiality and biaxiality of the liquid crystal.
To clarify the physical meaning of these quantities in $3D$, one may 
set the biaxiality parameter $\psi=0$ in Eq.~\eqref{eqn:diagQ}, to obtain a purely unixial profile and contract both Eqs.~\eqref{eqn:particle_nematic_tensor} and~\eqref{eqn:diagQ} with the tensor $\tilde{\mathbf{n}}\tilde{\mathbf{n}}$, with $\tilde{\mathbf{n}}=(1,0,0)$. One then obtains:
\begin{equation}
S(\mathbf{r},t) = \dfrac{1}{2} \langle 3 \cos^2 \theta - 1 \rangle,
\label{def:S}
\end{equation}
where $\theta(\mathbf{r})$ is the angle between $\tilde{\mathbf{n}}$ and the local orientation of the suspended particle $\bm{\nu}(\mathbf{r})$.
The scalar order parameter $S$ achieves its maximum in the perfectly aligned state (\emph{i.e.} $f_Q=\delta(\tilde{\mathbf{n}}-\bm{\nu})$) where $\langle  \cos^2 \theta \rangle=1$, while it falls to zero in the isotropic phase where the probability density $f_Q$ is uniform over the solid angle and $\langle  \cos^2 \theta \rangle=1/3$.
Therefore in the uniaxial limit ($\psi=0$) the Q-tensor admits the following vector representation
\begin{equation}
\mathbf{Q}= S\left( \mathbf{n}\mathbf{n} -\dfrac{1}{3} \mathbf{I} \right),
\label{eqn:Qtens_uniax}
\end{equation}
and all information on the order of the system are encoded in one scalar parameter $S$, addressing the degree of order, and a  director field $\mathbf{n}$ defining the alignment orientation.
However, in the most general case, both $S$ and $\psi$ are non null and fulfill the following relations~\cite{callan2006}:
\begin{align*}
\psi & \leqslant S \leqslant -3\psi +1, \\
0 &\leqslant \psi \leqslant 1/4.
\end{align*}
In this case, a full tensorial representation is necessary to completely address the orientational features of the liquid crystal.

\subsection{Active gel theory}
\label{sec:activegeltheory}
In the previous Section we have reviewed how the orientational features of a system with broken isotropic symmetry can be captured by a vector or tensor order parameter. These can be factually treated as slowly-varying variables thanks to the coarse-graining procedure, and as such they play the role of hydrodynamic variables in the active gel theory, in a similar fashion as in the continuum Landau-de Gennes theory for liquid crystals~\cite{degennes1993}.
A fundamental assumption of the active gel theory is the existence of a reference equilibrium state which defines the properties of the system in absence of internal energy injection. Moreover, one assumes that activity is only able to perturb the reference state in such a way that active systems never evolve \emph{too far} from equilibrium\footnote{For a proper discussion on how far from equilibrium active systems actually are, the interested reader can refer to \cite{Fodor2016}.} so that the evolution of the system can be derived in the framework of the Onsager theory for non-equilibrium thermodynamics, where a set of linear phenomenological relations between fluxes and forces can be established on the basis of symmetry considerations~\cite{Degroot2013}.
In the following, we will shortly review the Landau-de Gennes theory for \textcolor{black}{liquid crystals (LC)}, by writing the evolution equations for the order parameters and introducing a suitable free-energy which encodes the ordering features of a passive liquid crystal. Finally, we will consider how non-equilibrium effects due to the action of the active constituents can be incorporated in the theory through a phenomenological active stress in the momentum balance.

\paragraph{Landau-de Gennes free energy for passive \textcolor{black}{liquid crystals}.}
\label{sec:2b}

\begin{table*}[t]
\centering
\scriptsize 
\caption{The table summarizes bulk and elastic contributions to \textcolor{black}{the} free energy for polar and nematic, both uniaxial and biaxial, systems. Splay, twist and bending contributions have been written explicitly in terms of different elastic constants $K_i$ ($i=1,2,3$) for both polar and uniaxial nematic gels, while in the most general case of a biaxial nematic we did not distinguish between different contributions. 
Here, the three elastic constants $L_1,L_2,L_3$ are related to the Frank constant through the following relations $L_1 = (K_3+2K_2-K_1)/(9S^2), L_2 = 4(K_1-K_2)/(9S^2), L_3 = 2(K_3-K_1)/(9S^3)$ given that the Frank constants $K_i$ fullfill the condition $K_3 \geqslant K_1 \geqslant K_2$ to guarantee the positivity of $L_i$~\cite{schiele1983}.
The last line in the Table shows how the elastic contribution looks like assuming that the medium is elastically isotropic, \textit{i.e.}, $K_1=K_2=K_3=K$. Importantly, in the single constant approximation $L_2$ and $L_3$ become identically null and the only surviving elastic term is $\sim (\nabla \mathbf{Q})^2$.}
\label{tab:1}
\begin{tabular}{lllrc}
\toprule
Free energy contributions  &  & \multicolumn{1}{l|}{Polar Gel} & \multicolumn{2}{c}{Nematic Gel} \\ \cmidrule{4-5}
                                     &  & \multicolumn{1}{l|}{} & Uniaxial & Biaxial         \\ \cmidrule{1-5}
\multicolumn{2}{l}{Bulk}             & $a \mathbf{P}^2+b \mathbf{P}^4 $  & $r S^2 -wS^3 + u S^4 $ & $\tilde{r} Q_{ij}Q_{ji} -\tilde{w} Q_{ij}Q_{jk}Q_{ki} + \tilde{u} (Q_{ij}Q_{ji})^2 $                 \\ \hline

\multicolumn{1}{l|}{\multirow{3}{*}{Elastic}} & Splay & $\dfrac{K_1}{2} (\nabla \cdot \mathbf{P})^2 $  & \multicolumn{1}{l|}{$\dfrac{K_1}{2} (\nabla \cdot \mathbf{n})^2$}  & \multirow{3}{*}{} \\ \cmidrule{2-4}
\multicolumn{1}{l|}{}                  & Twist & $\dfrac{K_2}{2} (\mathbf{P} \cdot \nabla \times \mathbf{P})^2 $ &  \multicolumn{1}{l|}{$\dfrac{K_2}{2} (\mathbf{n} \cdot \nabla \times \mathbf{n})^2 $} & $ \dfrac{L_1}{2}(\partial_k Q_{ij})^2+ \dfrac{L_2}{2}(\partial_j Q_{ij})^2  + \dfrac{L_3}{2} Q_{ij} (\partial_i Q_{kl})(\partial_j Q_{kl}) $       \\ \cmidrule{2-4}
\multicolumn{1}{l|}{}                  & Bend & $\dfrac{K_3}{2} (\mathbf{P} \times \nabla \times \mathbf{P})^2 $ & \multicolumn{1}{l|}{$\dfrac{K_3}{2} (\mathbf{n} \times \nabla \times \mathbf{n})^2 $} &                   \\ \cmidrule{1-5}
\begin{tabular}[c]{@{}l@{}}Single constant\\ approximation\end{tabular} &  & $K(\nabla \mathbf{P})^2$ & $K(\nabla \mathbf{n})^2$ & $L_1 (\partial_k Q_{ij})^2$   \\ \bottomrule
\end{tabular}
\end{table*}

Macroscopic properties of the isotropic-nematic transition observed in lyotropic \textcolor{black}{liquid crystals} can be derived by a suitable free-energy functional $\mathcal{F}$ depending only on the order parameter of the system and its powers. This can be built on the base of symmetry considerations in the spirit of the Landau approach --\emph{i.e.} all terms appearing in the functional $\mathcal{F}$ must respect the symmetries of the disordered phase. In the case of LC, this means that only scalar terms invariant under rotations can be considered. Therefore, systems with polar symmetry described by a vector field $\mathbf{P}$ will admit terms of the kind $(\mathbf{P} \cdot \mathbf{P} )^m $, with $m$ a natural number. In the case of \textcolor{black}{systems} with nematic symmetry described by the $Q-$tensor instead, the only allowed terms are the traces of tensors which one can build out of $\mathbf{Q}$, hence: $Tr(\mathbf{Q}^2), Tr(\mathbf{Q}^3), Tr(\mathbf{Q}^4), Tr(\mathbf{Q}^2)^2,$ \emph{etc.}\footnote{Notice that no linear term is considered here since $Tr(\mathbf{Q})$ is null by definition. Moreover, in the case of nematic systems, odd powers can appear in the expansion as they respect isotropic symmetry. Conversely, for polar systems, these terms are not allowed as they do not transform trivially under space rotations.}
It must be stressed that the presence of a third order term in the Landau expansion is actually responsible for the first order nature of the nematic-isotropic transition characterized by a region of phase coexistence at the critical temperature between a liquid crystalline phase with birifrangent optical properties and the isotropic phase~\cite{chaikin1995}, while for polar system\textcolor{black}{s} the transition between the disordered and aligned state is second order. 
Table~\ref{tab:1} summarizes the bulk contributions to free energy for both polar and nematic systems; note that the uniaxial free energy can be derived from the biaxial case by writing the $Q-$tensor through Eq.~\eqref{eqn:Qtens_uniax}.

\begin{figure}[t]
\center
\resizebox{.95\columnwidth}{!}{%
  \includegraphics{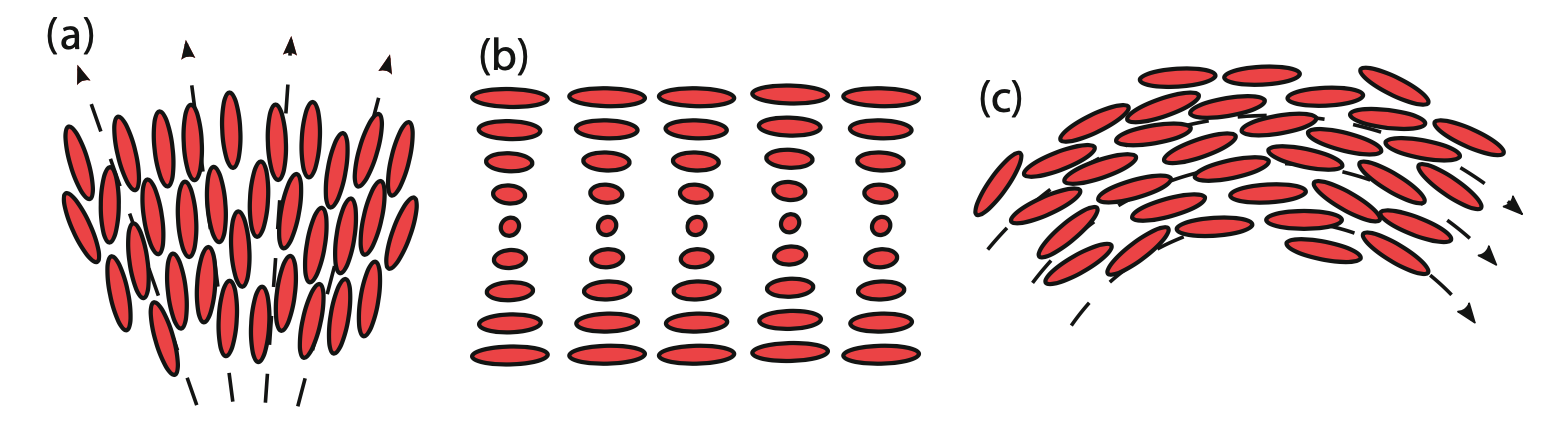}
}
\caption{ Cartoon of three modes of elastic deformations of a LC (red ellipsoids represent LC molecules): (a) Splay, (b) Twist and (c) Bend.}
\label{fig:2}       
\end{figure}

Another characteristic feature of LC relies on their elastic properties. 
Indeed, broken isotropic symmetry in liquid crystals implies that, while global rotations of the order parameter does not imply any energy cost, local deformations generating gradients of the order parameters, \textcolor{black}{do}. Three independent distortions can be identified in a nematic (or polar) LC, knowm as \emph{splay, twist} and \emph{bend} (see Fig.~\ref{fig:2} for a schematic representation).
Each of these is associated to a combination of gradients of the director field, corresponding to the following expression to be non-null:
\begin{align*}
\nabla \cdot \mathbf{n} \qquad \qquad \textit{Splay}, \\
\mathbf{n} \cdot (\nabla \times \mathbf{n}) \qquad \qquad \textit{Twist}, \\
\mathbf{n} \times (\nabla \times \mathbf{n}) \qquad \qquad \textit{Bend}. 
\end{align*}
The equilibrium configuration of nematic LC is the aligned state\textcolor{red}{\footnote{\textcolor{black}{More complex equilibrium configurations can be obtained by considering specific boundary conditions or geometrical confinement. This eventuality will be considered in detail in the next Sections.}}}, hence it is reasonable to associate an energy cost with each of these distortions\cite{Frank1958}, thus leading to the Frank free-energy\footnote{Analogous considerations can be done for polar LC. The elastic free energy in this case may be obtained from Eq.~\eqref{eqn:Frank_nem} through the formal substitution $\mathbf{n} \rightarrow \mathbf{P}$.}
\begin{equation}
\mathcal{F}^{Frank} = \dfrac{K_1}{2} (\nabla \cdot \mathbf{n})^2 + \dfrac{K_2}{2} (\mathbf{n} \cdot \nabla \times \mathbf{n})^2 +  \dfrac{K_3}{2} (\mathbf{n} \times \nabla \times \mathbf{n})^2,
\label{eqn:Frank_nem}
\end{equation}
where $K_1,K_2,K_3$ are three elastic constant respectively associated with splay, twist and bend distortions\footnote{Twist is forbidden in pure bidimensional systems, since this kind of deformation implies the director to coil around an axis, normal to the director itself.}.
However, in many situations it is convenient to adopt the \emph{single constant approximation}, by assuming the three constants to have the same value $K$. In this case, the Frank free energy assumes the simplified form
\begin{equation}
\mathcal{F}^{Frank} = \dfrac{K}{2} \left(\nabla \mathbf{n}\right)^2.
\end{equation}

The Frank theory is valid in the uniaxial approximation. However, this can be easily related to the Landau-De Gennes theory~\cite{degennes1993} which makes use of the most general form for the $Q$-tensor. The explicit expression of the elastic energy  is provided in Table~\ref{tab:1} which provides a complete summary of the free energy of both polar and nematic liquid crystals, including bulk and elastic contributions.

\paragraph{Evolution equations for Liquid Crystals.}
\label{sec:2e}

\begin{table*}[t]
\centering
\caption{Explicit expressions of the elastic (first row) and the interface (second row) stress, and of
the term $\mathbf{S}$ in the Equation~\eqref{eqn:generic_evolution_equation} (fourth row) for polar and nematic gels.
The molecular field $\bm{\Xi}$ is a vector, with components $h_{\alpha}$,  
for polar gels and a tensor $H_{\alpha \beta}$, for nematic gels,and its expression is shown in the third row. The flow-alignment parameters $\xi$ and $\xi'$ are respectively related to the polarization field $\mathbf{P}$ and to the nematic tensor $\mathbf{Q}$ and depend on the geometry of the microscopic constituents (for instance $\xi>0$, $\xi<0$ and $\xi=0$ for rod-like, disk-like and spherical particles, respectively). In addition, these parameters establish whether the fluid is flow aligning ($|\xi|>1$) or flow tumbling ($|\xi|<1$) under shear.}
\label{terms}
\resizebox{\textwidth}{!}{
\begin{tabular}{l|l|l}
\toprule
                                   & Polar Gel                                                                                                                                                                 & Nematic Gel                                                                                                                                                                                                                                                                                                                                                                                                                                  \\
\midrule
\rule{0pt}{6ex}
$\sigma_{\alpha\beta}^{el}$   & $\frac{1}{2}(P_{\alpha}h_{\beta}-P_{\beta}h_{\alpha})-\frac{\xi}{2}(P_{\alpha}h_{\beta}+P_{\beta}h_{\alpha})-K\partial_{\alpha}P_{\gamma}\partial_{\beta}P_{\gamma}$ & \begin{tabular}[c]{@{}l@{}} $2\xi'\left(Q_{\alpha \beta}-\frac{\delta_{\alpha \beta}}{3}\right)Q_{\gamma \nu}H_{\gamma \nu}-\xi' H_{\alpha \gamma}\left(Q_{\gamma \beta}+\frac{\delta_{\gamma \beta}}{3}\right)$\\$-\xi'\left(Q_{\alpha \gamma}+\frac{\delta_{\alpha\gamma}}{3}\right)H_{\gamma\beta}-\partial_{\alpha}Q_{\gamma\nu}\frac{\delta\mathcal{F}}{\delta\partial_{\beta}Q_{\nu\gamma}}+Q_{\alpha\gamma}H_{\gamma\beta}-H_{\alpha\gamma}Q_{\gamma\beta}$\end{tabular} \\
\rule{0pt}{5ex}

$\sigma_{\alpha\beta}^{int}$ & $\left( f-\phi\frac{\delta \mathcal{F}}{\delta\phi} \right)\delta_{\alpha\beta} - \frac{\partial f}{\partial\left(\partial_{\beta}\phi\right)} \partial_{\alpha}\phi$               &  $\left( f-\phi\frac{\delta \mathcal{F}}{\delta\phi} \right)\delta_{\alpha\beta} - \frac{\partial f}{\partial\left(\partial_{\beta}\phi\right)} \partial_{\alpha}\phi$                                                                                                                                                                                                                                                                                                                                                                                                                                             \\
\rule{0pt}{5ex}

$\sigma_{\alpha\beta}^{act}$ & $-\zeta \left( P_\alpha P_\beta -\dfrac{1}{3} P^2 \delta_{\alpha \beta} \right) +\bar{\zeta} \epsilon_{\alpha\beta\mu}\partial_{\nu}\phi P_{\mu} P_{\nu}$  &  $-\zeta Q_{\alpha \beta} + \bar{\zeta} \epsilon_{\alpha\beta\mu}\partial_{\nu}\phi Q_{\mu \nu}$                                                                                                                                                                                                                                                                                                                                                                                                                                             \\
\rule{0pt}{5ex}

$\bm{\Xi}$ & $  h_{\alpha} = -\dfrac{\delta \mathcal{F}}{\delta P_\alpha} $ & $  H_{\alpha \beta} = -\dfrac{\delta \mathcal{F}}{\delta Q_{\alpha \beta}} + \left( \dfrac{\delta \mathcal{F}}{\delta Q_{\gamma \gamma}} \right) \delta_{\alpha \beta}$                                                                                                                                                                                                                                                                                                                                                                                                                                             \\
\rule{0pt}{5ex}

$\mathbf{S}$ & $ -\Omega_{\alpha \beta} P_{\beta} +\xi D_{\alpha \beta} P_{\beta}$                                                                              & \begin{tabular}[c]{@{}l@{}} $ [\xi' D_{\alpha \gamma} +\Omega_{\alpha \gamma}]\left(Q_{\gamma \beta}+\frac{\delta_{\gamma \beta}}{3}\right)+\left(Q_{\alpha \gamma}+\frac{\delta_{\alpha \gamma}}{3}\right)[\xi' D_{\gamma \beta} -\Omega_{\gamma \beta}]$\\$-2\xi'\left(Q_{\alpha \beta}+\frac{\delta_{\alpha \beta}}{3}\right)\left(Q_{\gamma \delta} \ \partial_{\gamma} v_{\delta} \right)$ \end{tabular}    \\

\bottomrule
\end{tabular}
}
\end{table*}

The evolution of a liquid-crystalline systems mostly depends upon two separate mechanisms: on the one hand the LC tends to evolve towards the equilibrium configuration which minimize the free-energy functional $\mathcal{F}$, on the other hand LC molecules are advected by the flow $\mathbf{v}(\mathbf{r},t)$. This may be in turn externally excited, as in the case of a shear or Poiseuille flow, or generated by stresses arising due to the internal structure of the system (mostly gradients of the orientational order parameters). In this latter case it is customary to speak of \emph{backflow}.

Regardless of the different tensor nature of the fields introduced in the previous section, the generic order parameter $\bm{\Psi}(\mathbf{r},t)$ (which we use here to denote either by $\mathbf{P}$ or $\mathbf{Q}$ indifferently) evolves according to an advection-relaxation equation of the kind
\begin{equation}
(\partial_t +   \mathbf{v} \cdot \nabla) \bm{\Psi} - \mathbf{S} = \mathcal{G}  \bm{\Xi} (\bm{\Psi}),
\label{eqn:generic_evolution_equation}
\end{equation}
known as \emph{Beris-Edwards equation}~\cite{beris1994}, where $\mathbf{v}(\mathbf{r},t)$ denotes the flow velocity. The right-hand side of Eq.~\eqref{eqn:generic_evolution_equation} is a thermodynamic force which drives the relaxation of $\bm{\Psi}$ towards its equilibrium state.
This is given by the molecular field $\bm{\Xi}$ with the same tensor features as the associated dynamical field, and it is derivable from the free-energy as the functional derivative of $\mathcal{F}$ with respect to $\bm{\Psi}$, as indicated in the third row of Table~\ref{terms}.
Moreover, this is multiplied by a phenomenological constant $\mathcal{G}$ which defines the relevance of relaxation with respect to advection.
This is in turn accounted by the differential operator on the left-hand side of Eq.~\eqref{eqn:generic_evolution_equation}, \emph{i.e.} the material derivative $d_t= (\partial_t +   \mathbf{v} \cdot \nabla)$, which provides a description of the  transport of the order parameter due to fluid advection. Aligning hydrodynamic interactions are incorporated into the strain-rotational derivative $\mathbf{S}=\mathbf{S}(\nabla \mathbf{v}, \bm{\Psi})$, whose explicit expression depends on the tensor nature of the order parameter (see Table~\ref{terms}).
This can be formally derived making use of Liouville equations, since for Hamiltonian system one can write $D_t\bm{\Psi}=\partial_t{\bm{\Psi}}+ \lbrace \bm{\Psi}, \mathcal{H}\rbrace$, where $\lbrace...\rbrace$ are the Poisson brackets and the Hamiltonian is $\mathcal{H} = \mathcal{F} + \frac{1}{2} \int \rho \mathbf{v}^2 $.

A more phenomenological procedure to derive the material derivative explicitly is based on the fact that the order parameter can be transported, rotated or stretched due to fluid advection. In the case of a polarization field, for instance, the relative position $\tilde{\mathbf{r}}$ of two close points in the fluid evolves according to the following equation:
\begin{equation}
D_t \tilde{\mathbf{r}} = \partial_t \tilde{\mathbf{r}} + (\mathbf{v} \cdot \nabla) \mathbf{v} + \bm{\Omega}\cdot \tilde{\mathbf{r}} + 
 \cdot \mathbf{D} \cdot \tilde{\mathbf{r}} ,
\label{eqn:material_derivative_reconstruction}
\end{equation}
where $\mathbf{D}= \frac{1}{2}(\nabla \mathbf{v} + \nabla \mathbf{v}^T)$  and $\bm{\Omega}= \frac{1}{2}(\nabla \mathbf{v} - \nabla \mathbf{v}^T)$
respectively denote the symmetric and anti-symmetric part of the gradient of the velocity field $\mathbf{W}=\nabla \mathbf{v}$.
The first two contributions are the usual terms which appear in the Lagrangian derivative, while the third and fourth ones account respectively for rigid rotations and stretching of the fluid element, respectively. Therefore, the material derivative for the polarization field will include the first three terms since a vector advected by the flow is capable to follow any rigid motion. The last term in Eq.~\eqref{eqn:material_derivative_reconstruction}, cannot directly enter the material derivative of a vector field, but it must be weighted through an alignment parameter $\xi$, which rules the dynamical behavior of the vector field under enlargement and/or tightening of flow tubes. This allows us to obtain the material derivative for the polarization field simply substituting $\mathbf{P}$ in place of $\tilde{\mathbf{r}}$.

\paragraph{Advection-diffusion equation for scalar concentration fields.}
In many experiments on active gels (and more generally on complex fluids) the total number of suspended particles can be assumed constant in time, but they can still arrange in complex patterns in space, for instance to arrange in vesicles or droplets. Even in this case one can assume that the equilibrium state is defined by a suitable free-energy $\mathcal{F}=\mathcal{F}[\phi,\dots]$, whose expression explicitly depends on the field configuration and whose minimization provides the equilibrium state of the system. In the case of a scalar order parameter, Eq.~\eqref{eqn:generic_evolution_equation} can be written as a Cahn-Hilliard equation where the term $\bm{\Xi}$ is written as
$$
\bm{\Xi}= \nabla \cdot \mathbf{j}
$$ 
being $\mathbf{j}$ a diffusive flux for $\phi$ that can be defined as the gradient of a thermodynamic force $\mu=\mu(\mathbf{r},t)$ which vanishes at equilibrium. A standard  choice for the thermodynamic force $\mu(\mathbf{r},t)$ is given by
\begin{equation}
\mu (\mathbf{r},t) = \dfrac{\delta\mathcal{F}[\phi,\dots]}{\delta \phi}(\mathbf{r},t),
\label{eqn:def_chemical_potential}
\end{equation}
where $\delta/\delta \phi$ denotes the functional derivative with respect to the concentration field, so that the Cahn-Hilliard equation can be written as
\begin{equation}
\partial_t \phi + \nabla \cdot (\phi \mathbf{v}) = M \nabla^2 \mu,
\label{eqn:Cahn_Hilliard}
\end{equation}
where $M$ is the mobility of the diffusive field. Importantly, the only allowed hydrodynamic coupling is the advective term $\nabla \cdot (\phi \mathbf{v})$ since the strain-rotational derivative $\bm{S}$ is null due to the scalar nature of the concentration field.

\paragraph{Hydrodynamic equations.}

Finally, we will introduce the hydrodynamic equations for (active) liquid crystals. Evolution equations for mass density $\rho (\mathbf{r},t)$ and velocity $\mathbf{v}(\mathbf{r},t)$ are respectively given by
\begin{eqnarray}
\partial_t \rho + \nabla \cdot ( \rho \mathbf{v}) & = & 0, \label{eqn:cont_eq}\\
\rho\left(\partial_t+\mathbf{v}\cdot\nabla\right)\mathbf{v} & = & -\nabla p + \nabla\cdot\bm{\sigma}. \label{eqn:nav}
\end{eqnarray}
Eq.~\eqref{eqn:cont_eq} is the continuity equation for mass density. 
In most systems where droplets exhibit self-propulsive features the typical Mach number $Ma$ of the flow, defined as the ratio of the stream velocity and the speed of sound, is small; in such limit, the continuity equation simplifies to the solenoidal condition for the velocity field
\begin{equation}
\nabla \cdot \mathbf{v}=0 + \mathcal{O}(Ma^2),
\label{cont}
\end{equation}
so that the fluid in this regime can be assumed at all practical effects as incompressible.
Eq.~\eqref{eqn:nav} is the Navier-Stokes equation, where $p$ is the ideal fluid pressure
 and $\bm{\sigma}$ is the stress tensor~\cite{beris1994} that we can ideally split into a reactive/passive and a non-equilibrium/active contribution:
\begin{equation}
\bm{\sigma}=\bm{\sigma}^{\textit{pass}}+\bm{\sigma}^{\textit{act}}.
\end{equation}
The passive part is, in turn, the sum of three terms:
\begin{equation}
\bm{\sigma}^{\textit{pass}}=\bm{\sigma}^{\textit{visc}}+\bm{\sigma}^{\textit{el}}+\bm{\sigma}^{\textit{int}}.
\end{equation}
The first term is the viscous stress, which can be written as
\begin{equation}
\sigma_{\alpha\beta}^{visc}=\eta(\partial_{\alpha}\mathrm{v}_{\beta}+\partial_{\beta}\mathrm{v}_{\alpha}),
\label{eqn:viscous}
\end{equation}  
where $\eta$ is the shear viscosity.
This term is responsible for dissipation of energy due to molecular dissipative interactions.
Moreover, elastic deformations of the liquid crystal (either polar or nematic) as well as interfaces in the concentration field generate local stresses,  $\bm{\sigma}^{\textit{el}}$ and $\bm{\sigma}^{\textit{int}}$, responsible for backflows. These have the effect of fastening relaxation, in absence of any internal or external forcing. They are strongly related to the equilibrium properties of the order parameters and for this reason are customarily addressed as \emph{reactive}.
They can be derived by either looking at the symmetries through the Onsager theory of fluxes and forces or by a Hamiltonian approach, through the Noether theorem, in which case the stress tensor would play the role of the momentum conserved current. The expression for the elastic and interface stress in terms of the generic free energy $\mathcal{F}$ is reported in Table~\ref{terms}.

\subsection{Active Forces and intrinsic instability of active gels in 2D}

\begin{figure}[t]
\center
\resizebox{.8\columnwidth}{!}{%
  \includegraphics{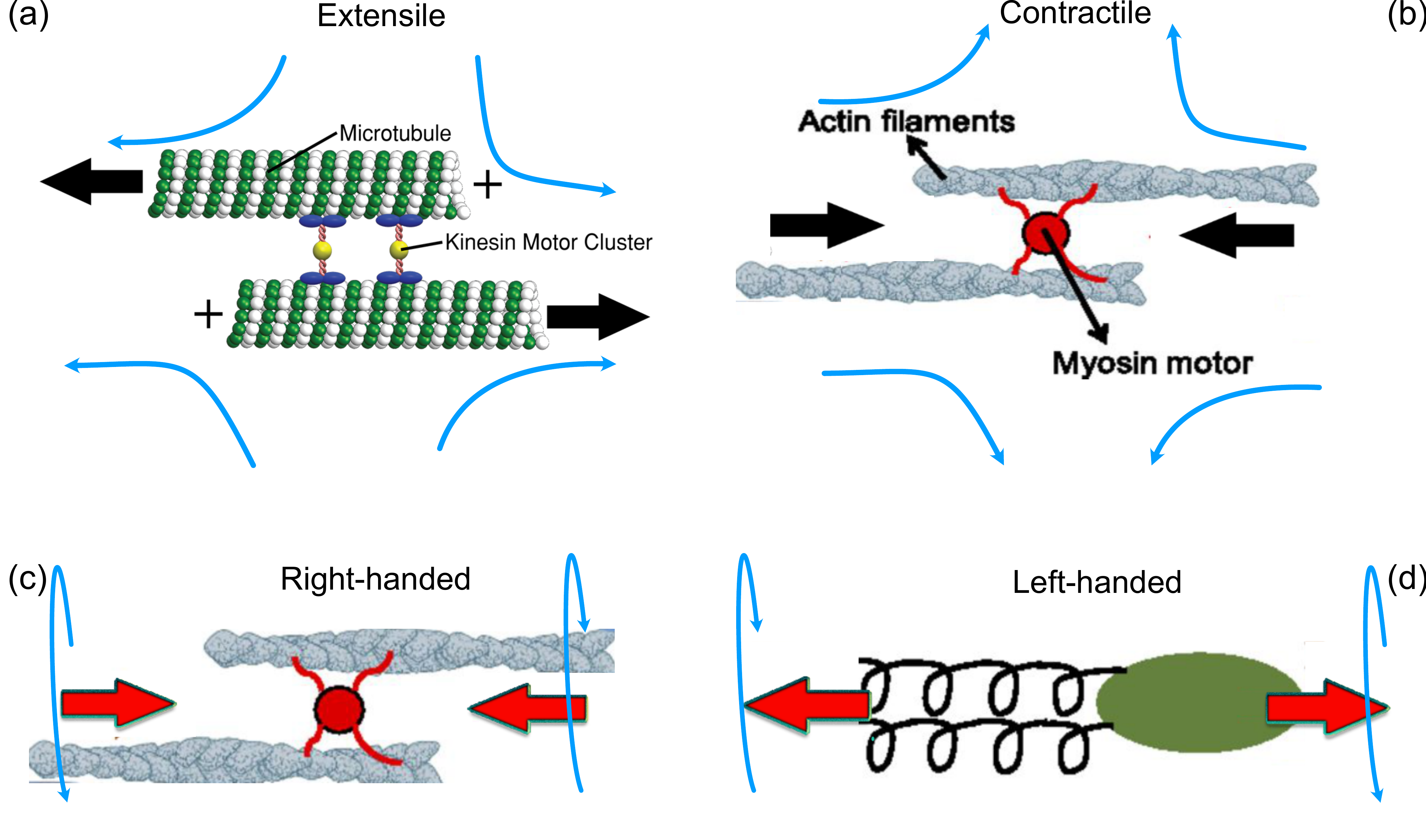}
}
\caption{ \textbf{Flow generated by active agents.} In the case of MT bundles (a), fluid is expelled both forwards and backwards along the fore-aft axis (cyan arrows) as the result of the extension of the fibers (black arrows), and drawn inwards radially towards their axis, generating an extensile flow pattern.
Conversely, the motor proteins which activate actomyosin fibers (b) pull the filaments, leading to the contraction of the system. This results into a contractile flow. Many biological microswimmers may exert on the surrounding fluid not only force dipoles but also torque dipoles. Actin filaments that make up the actomyosin network are twisted into right-handed helices (c) whereas the flagella in bacterial self-propulsion (d) are left-handed helices. \textcolor{black}{Panels (b)-(d) adapted from Ref.~\cite{Tjhung2017}.}}
\label{fig:3}       
\end{figure}

The theory presented until now describes the nemato-hydrodynamics of passive gels where all contributions to the stress tensor and all terms appearing in the evolution equations can be derived from the free energy functional defining the equilibrium properties of the system. 
In this Section we will focus on how the theory can be adapted to consider internal energy injection due to the self-motile individual constituents. To begin with, we stress the fact that the usage of the term \emph{active} in the context of a continuum approach relies on the fact that the origin of the extra non-equilibrium terms which one needs to introduce is purely phenomenological as it cannot be derived from first principles.
A  first form for an active stress $\bm{\sigma}^{act}$ was proposed in~\cite{pedley1990}  and later developed in the context of a coarse-grained model for active filaments or orientable particles~\cite{simha2002,kruse2004,julicher2007}. 
In a seminal work of 2002, Simha and Ramaswamy~\cite{simha2002} produced a formal derivation of the active stress tensor for active nematics, which in the last two decades has somehow become a well established paradigm in the scope of active matter. In this Chapter we will not reproduce the analytical argument of~\cite{simha2002}, but the interested reader can refer to the Chapter of J. Yeomans where Simha and Ramaswamy's derivation has been explicitly reviewed.

The form of the active stress for an active nematics can be expressed in terms of the Q-tensor as
\begin{equation}
\bm{\sigma}^{act} = -\zeta \phi \mathbf{Q}.
\label{eqn:active_stress_tensor_Q}
\end{equation}
where $\zeta$ is a phenomenological parameter which measures the strength of active forcing exerted by the active particles on the surrounding environment. Moreover, the expression for the active stress tensor for a system of polar (rather than nematic) particles can be otained from Eq.~\eqref{eqn:active_stress_tensor_Q} by taking the uniaxial approximation for the Q-tensor and identifying the director field $\mathbf{n}$ with the polarization field, thus obtaining
\begin{equation}
\bm{\sigma}^{act} = -\zeta \phi \left( \mathbf{P} \mathbf{P} - \dfrac{\mathbf{P}^2}{d} \mathbf{I}\right),
\label{eqn:active_stress_tensor_P}
\end{equation}
being $d$ the dimensionality of the system.

Before proceeding with some quantitative arguments on the effects that such an active term has on the nemato-hydrodynamics of the system, we want to draw the attention of the reader on a couple of points.
First, we observe that the physical quantity entering the Navier-Stokes equation~\eqref{eqn:nav} is the divergence of the stress tensor representing a factual active force $\mathbf{f}^{act} \sim \nabla \cdot \bm{\sigma}^{act}$, rather than the stress tensor itself. Therefore, one can easily guess that this term would play an important role when the LC pattern achieves non-uniform patterns which would eventually lead to Q-tensor gradients. Conversely, in a perfectly aligned state where $\mathbf{Q}$ is uniform, $\mathbf{f}^{act}=0$, despite the active stress would attain a non-null (but uniform) profile.
The second observation concerns the physical meaning of the sign of the activity parameter $\zeta$.
In general, the propulsive motion of active particles in a fluidic environment produces a circulating flow pattern in the far field whose direction mostly depends on the mechanism of propulsion implemented by the swimmer. For instance, in the case of MT bundles, fluid is expelled both forwards and backwards along the fore-aft axis as the result of the extension of the fibers themselves, and drawn inwards radially towards their axis, generating an \textit{extensile} flow pattern (see Fig.~\ref{fig:3}(a)). This behavior corresponds to positive values of the activity parameter $\zeta>0$. Conversely, the motor proteins which activate actomyosin fibers pull the filaments, leading to the contraction of the system. This results into a \textit{contractile} flow, opposite to that of the previous example, as shown in Fig.~\ref{fig:3}(b), so that in this case, the activity parameter would be negative, $\zeta<0$.

To conclude, we briefly mention here that many biological microswimmers (from acto-myosin bundles to bacteria) may exert on the surrounding fluid not only force dipoles but also 
torque dipoles~\cite{Zhou2014,naganathan2014,livolant1986,livolant1991}, which result in a source of angular momentum.
\textcolor{black}{For instance, the actin filaments that make up the actomyosin network are twisted into right-handed helices (Fig.~\ref{fig:3}(c)), whereas the flagella in bacterial self-propulsion are left-handed helices (Fig.~\ref{fig:3}(d)).}
If the net torque is null but torque dipoles do not vanish, a coarse-graining procedure ~\cite{furthauer2012,Tjhung2017,maitra2019} analogous to that of Simha and Ramaswamy, shows that a suitable choice for the nematic chiral stress tensor is given by $\bar{\zeta} \epsilon_{\alpha\beta\mu}\partial_{\nu}\phi Q_{\mu \nu}$  with $\epsilon_{\alpha \beta \mu}$ the third order Levi-Civita tensor. The physical meaning of the phenomenological constant $\bar{\zeta}$ is not as straight-forward as its force-dipole analogue and a more appropriate discussion of its effect on the dynamics of a (chiral) liquid crystal is postponed to Section~\ref{sec:chol}, where we will consider how chiral LC droplets can be implemented to produce controllable propulsive states.

\paragraph{Instability of active gels.}

In this Section we will review some topical results in the field of the active gel theory, \emph{i.e.} how activity and elasticity cooperate to give rise to self-sustained and stationary flows.

\begin{figure}[ht!]
\center
\resizebox{.6\columnwidth}{!}{%
  \includegraphics{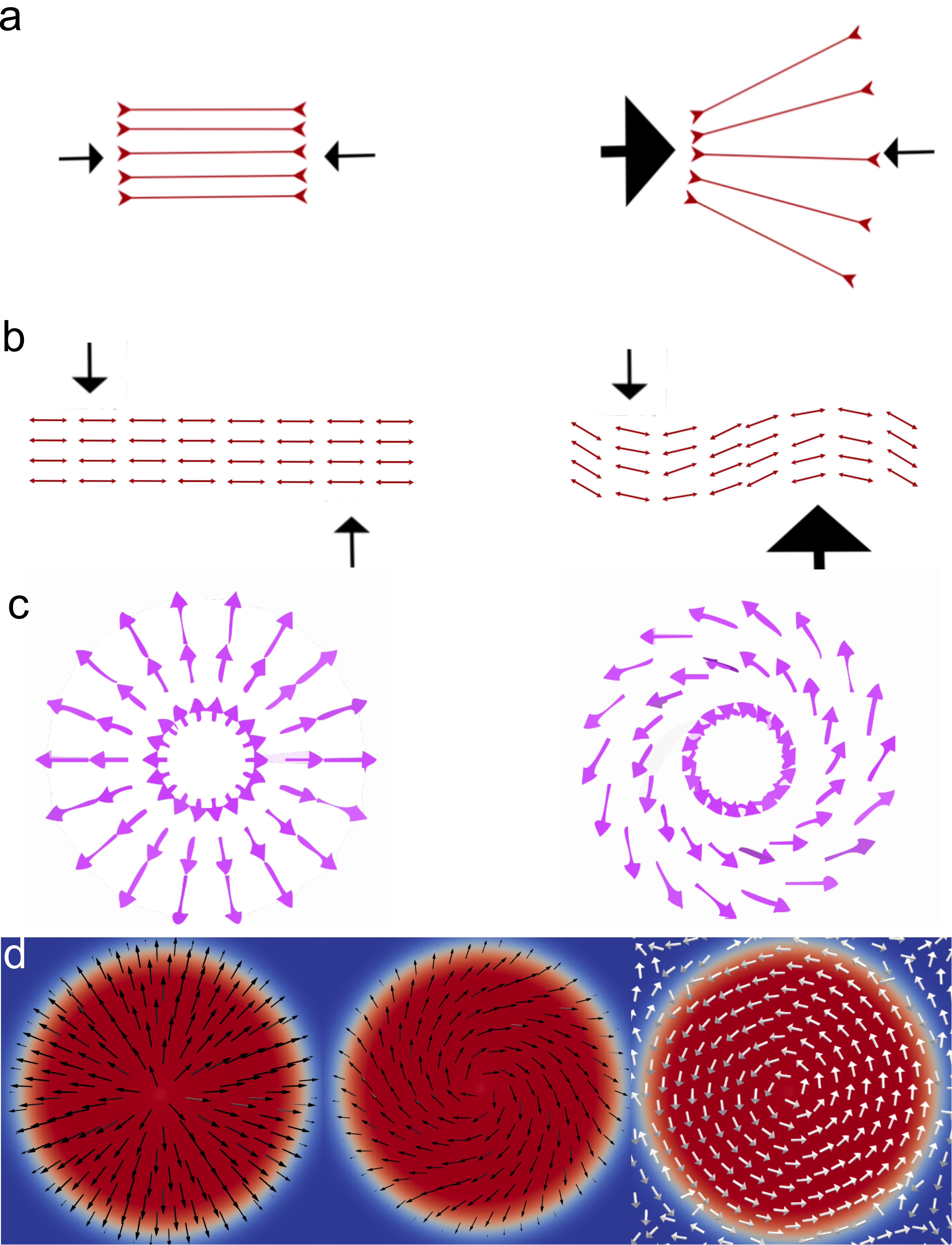}
}
\caption{ \textbf{Instability of active gels.} \textbf{(a)} Sketch of instability and spontaneous symmetry breaking mechanism for contractile systems. When the system is completely ordered (left panel) force dipoles compensate each other, while if a splay deformation is present  the density of contractile forces is greater on the left than on the right. This determines a flow that produces further splay (right panel), resulting in a macroscopic flowing state. \textbf{(b)} Extensile fluids instead get unstable to bend deformations. \textbf{(c)}  Schematic representation of the geometry of point defects of topological charge one in two dimensions. Displayed are the orientations of polarization vectors as purple arrows for  aster (left) and spiral (right).
\textbf{(d)} Results from numerical simulations of an extensile polar droplet with homeotropic hanchoring, obtained using  the model presented in Section 1.3.2. Black arrows represent the polarization field, while white ones represent the velocity field. At vanishing activity (left panel) the droplet attains a spherical shape and the liquid crystalline pattern arranges into an aster.
As activity exceeds a critical threshold bending instability lead to a spiral pattern (central panel) while the velocity field develops a vortex with its center in correspondence of the topological defect of the polarization (right panel).}
\label{fig:4}       
\end{figure}

\textcolor{black}{The continuum theory described in the previous Section exhibits an intrinsic instability due to the active injection of energy. 
Importantly, active gels exhibit different kinds of instabilities according to the kind of activity: indeed while extensile gels are unstable to bend deformations, contractile gels are unstable to splay~\cite{giomi2008,edwards2009}.
We will here provide a phenomenological explanation of the instability. The interested reader may refer to the Chapter of J. Yeomans for a quantitative analysis of this topic.} \textcolor{black}{We consider, for instance, the case of contractile dipoles initially in a perfectly aligned state, as sketched on the left of Fig.~\ref{fig:4}. The force dipoles are here balancing each other and the net flow is null. However, in presence of a small splay deformation, the density of contractile forces on the left is larger than the one on the
right, and a flow sets up. Such a flow, directed from left to right in panel~(a), destabilizes the system by strengthening the initial splay and the perturbation grows larger and larger leading to a macroscopic flowing state. 
Conversely in an extensile system, configuration of panel~(a) would generate a flow directed from the right to the left which restores the initially aligned state so that splay deformations are rapidly dumped out.
However, extensile systems are unstable to bending deformations, as can be explained by a similar argument.
This situation is sketched in panel~(b) of a Fig.~\ref{fig:4}.}

\textcolor{black}{How does the intrinsic instability of active gels relate to the motile properties of active droplets? To answer this question}
we consider here the simple case of a liquid crystal with polar symmetry confined in a 2D flat circular domain of radius $R$. For the sake of simplicity we require the polarization field to be a unit vector ($|\mathbf{P}|=1$), and both the polarization and the velocity field $\mathbf{v}$ to be axisymmetric --hence they only depend on the radial distance from the center of the disk $r \le R$ but not on the angular position $\varphi$ ($\mathbf{P}=\mathbf{P}(r)$ and $\mathbf{v}=\mathbf{v}(r)$).
Moreover, the velocity field satisfies no-slip boundary conditions, which implies that all components of velocity field identically vanish at the boundary.

At equilibrium (\emph{i.e.} in absence of activity) the liquid crystal features an aster at the center of the disk (see left panel of Fig. \ref{fig:4}(c)) in absence of any flow ($\mathbf{v}=0$). This is the simplest example of topological defect, namely a region of space (in this case just a point) where the order parameter cannot be unequivocally defined. 
However, the nemato-hydrodynamic equations presented in the previous Section exhibit an instability to active injection. As said,   this means that the energy provided by the activity is both used to deform the liquid crystalline pattern and sustain internally generated flows. Importantly, cylindrical symmetry and incompressibility yield  the radial component of the velocity field $v^{r}=0$, so that $\mathbf{v}= \frac{1}{r} v^{\varphi}\mathbf{e}_{\varphi}$, with $v^{\varphi}=v^{\varphi}(r)$ and $(r,\varphi)$ polar coordinates with associated basis of normal unit vectors $(\mathbf{e}_{r}, \mathbf{e}_{\varphi})$.
With this choice of notation, the polarization field can be written as $\mathbf{P} = \cos\theta\,\mathbf{e}_{r}+\frac{1}{r}\sin\theta \mathbf{e}_{\varphi}$ with $\theta$ a constant. 

Solving the Navier-Stokes equation Eq.~\eqref{eqn:nav} and the Ericksen-Leslie equation \eqref{eqn:generic_evolution_equation} for a flow-aligning liquid crystal ($|\xi|>1$) one finds after standard algebraic manipulations (see e.g. Ref. \cite{kruse2004, Giomi2014, Hoffmann2021}):
\begin{align}
\theta = \pm \frac{1}{2} \arccos(-1/\xi) \;, \label{eq:sol_1}\\
v^{\varphi} = \pm \frac{\zeta}{\eta}\,\frac{\sqrt{1-1/\xi^{2}}}{2}\,r\log\frac{r}{R}\;,\label{eq:sol_2}\\
p = \frac{\zeta}{\xi} \log \frac{r}{R} - 2\pi K \delta(\mathbf{r})\; \label{eq:sol_3},
\end{align}
with $\delta(\cdot)$ the delta-function on the disk and $K$ the elastic constant of the polar liquid crystal in the single elastic constant approximation. \textcolor{black}{Therefore, the resulting non-equilibrium steady state is characterized by a spiral in the pattern of the polarization field, as shown in the right panel of Fig.~\ref{fig:5}(c), and a vortical flow field.}
Finally, by looking at Eq. (\ref{eq:sol_3}), we observe that  the pressure diverges at the origin, in correspondence of the $+1$ topological defect. This divergence stems from the continuous description of defects. However, this can be regularized, as customary in these context, by considering a core radius $r_{\rm c}$. This can be done, for instance, by setting a short length scale cut-off, so that $r_{\rm c}\le r \le R$.

\textcolor{black}{The arguments presented in this Section show that motile states are obtained in the context of the active gel theory as the result of the mutual competition of various intrinsic properties of the active fluid, \emph{i.e.} rate of active energy injection, viscosity, elasticity, but also of the geometry of the system, \emph{i.e.} typical confining size. This becomes especially relevant in the context of  active liquid crystalline droplet\textcolor{red}{s}. Indeed,  in the following Sections we will discuss how the possibility of achieving such self-sustained  flowing states through the confinement of the liquid crystal and the resulting topological constraint which are often at the base of most motile properties of active droplets and shell both in 2 and 3-dimensional realizations.}

\subsection{Numerical results of active droplets in $2D$}
\label{sec:3.4}
The results of the previous Section provide a clear example of how activity and confinement might lead to non-trivial flow structures with important motility properties. This is a crucial topic, since many biological realizations of active fluids are actually found in confined environments (\emph{e.g.} cells, vesicles, bacterial colonies, \emph{etc} ). In this respect, we will now consider the case of active droplets, where an active liquid crystal is confined in a droplet embedded in an isotropic background. This system is particularly appealing as it can be used as a minimal model for a number of experimental and biological realizations. 

Before getting involved into the discussion of the mechanical properties of active droplets, we want to draw the attention of the reader on the role of the anchoring of the liquid crystal. Indeed, in the following we will show that most dynamical states of active droplets mostly depend on two parameters: the activity (either contractile or extensile) and the anchoring. The latter defines the equilibrium orientation of the liquid crystal at the droplet interface and it can either be hometropic, if the preferential direction is perpendicular, or tangential. 
These two different kinds of anchoring also lead to different equilibrium configurations (in absence of activity). Indeed, while homeotropic anchoring leads to the development of an aster at the center of the droplet, as in the previous section, tangential anchoring results into a vortex. These two defects are only apparently different, but from a topological perspective they both hold the same topological charge ($+1$) as they both perform a rotation of $2\pi$ around the defect core. However, the pattern resulting from the configuration may lead to very different dynamical properties when activity is considered. Indeed, while extensile activity is able to excite bending instability (but not splay ones) the opposite occurs in the case of contractile active stresses. In this case, bending perturbations are suppressed, while splay deformations can grow larger and larger. 

To include anchoring in the continuous modelling, we observe that from a numerical perspective an active droplet can be described in terms of the concentration $\phi$ of active material~\cite{bone2017,negro2018}. In many practical situation this can be assumed as a conserved, phase separating field whose bulk free energy density is given by a standard $\phi^4$ theory:
\begin{equation}
f^{bulk} = -\dfrac{a}{2} \phi^2 + \dfrac{a}{4} \phi^4 + \dfrac{k_\phi}{2} (\nabla \phi)^2.
\label{eqn:phi4}
\end{equation}
Here, $a, k_\phi>0$ are two phenomenological constants defining the surface tension $\Sigma=\sqrt{8Ka/9}$ of the droplet. 
Anchoring can be easily introduced in the model by considering a boundary term in the free energy, coupling the order parameter to the gradients of the concentration field $\phi$, which in this context naturally defines the droplet itself. A suitable free-enegy density term is given by
\begin{equation}
f^{anch} = 
\begin{cases}
W (\mathbf{P}\cdot \nabla \phi)^2 \qquad & \text{for polar systems,}\\
W (\nabla \phi \cdot \mathbf{Q} \cdot \nabla \phi)\qquad & \text{for nematic systems.}
\end{cases}
\end{equation}
Here $W$ is the anchoring parameter (or simply anchoring) and its sign defines the kind of anchoring. Therefore, energy is minimized if $\mathbf{P} \perp \nabla \phi$ when $W>0$ resulting in a configuration with the polarization parallel to the interface. Conversely, when $W<0$ minimization of $f^{anch}$ requires $\mathbf{P}\cdot \nabla \phi$ to be as large as possible, favoring a homeotropic configuration.

\textcolor{black}{To compare the numerical result with the anaylitical case of the previous Section, we begin by considering the case with extensile activity and homeotropic anchoring.} At vanishing activity the droplet is practically passive: it attains a spherical shape and the liquid crystalline pattern arranges into an aster, as prescribed by the homeotropic anchoring (see for instance left panel  of Fig. \ref{fig:4} (d)). \textcolor{black}{As activity is increased, we expect the active liquid crystal to undergo a process similar to the bending instability described in the previous Section. Indeed, as $\zeta$ exceeds a critical threshold (which is inversely proportional to the radius of the droplet) the polarization field settles into a spiral pattern} (see central panel of Fig. \ref{fig:4} (d)) while the velocity field develops a vortex  as shown in the right panel of Fig. \ref{fig:4} (d) and predicted by the analytical argument presented before. 

This result suggests the important conclusion that rotational dynamics results from the breaking of mirror symmetry leading to the production of angular momentum. Reasoning along similar lines, one may infer that translational dynamics can be obtained by breaking polarity inversion symmetry\footnote{\textcolor{black}{Configurations which are symmetric with respect to polarity inversion do not change by flipping the vectors in the whole configuration.}}, which would subsequently lead to the production of linear momentum. However, strong anchoring conditions prevent to break inversion symmetry. Therefore, in the next section we will consider the case of weaker anchoring to study how linear motility can arise in active droplets.

\begin{figure}[ht!]
\center
\resizebox{.7\columnwidth}{!}{%
  \includegraphics{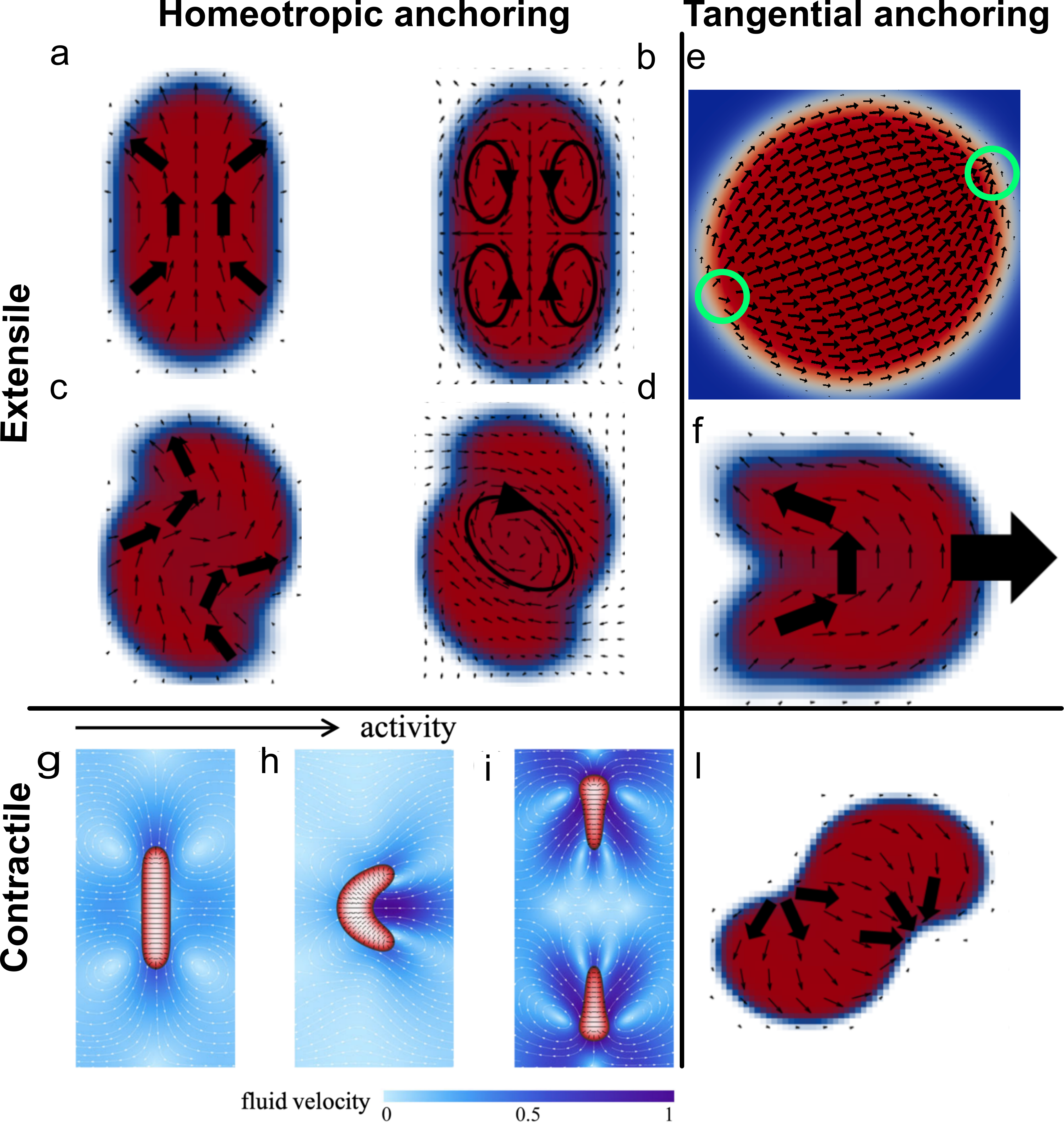}
}
\caption{ \textbf{Rotational and translational motion of 2D active droplets.} Configuration of the polarization field in the \textbf{(a)} quiescent and \textbf{(c)} rotational regime for a contractile droplet with homeotropic anchoring. Panels (b) and (d) show the the velocity field for the two cases. For extensile activity with tangential anchoring the quiescent state (e) is characterized by two singular points at the droplet surface (green circles) where anchoring is not satisfied. Enlarging activity in this case causes the singular points at droplet interface to rearrange at the rear (panel (f)). Such a configuration is not axisymmetric and the liquid crystal deformation produces a net flow in the direction normal to the bending. \textbf{(g)-(i)} Director field (black rods) and flow configurations (color plot with superimposed flow stream lines) of a contractile nematic droplet with homeotropic anchoring for different values of activity~\cite{Giomi2014}. \textbf{(l)} Polarization field configuration (black arrows) of a polar contractile droplet with tangetial anchoring. \textcolor{black}{Panels (a-d), (f) and (l) adapted from Ref.~\cite{fialho2017} with permission from the Royal Society of Chemistry. Panels (g-i) adapted from Ref.~\cite{Giomi2014} with permission from L. Giomi, Copyright 2014 American Physical Society.}}
\label{fig:5}       
\end{figure}

\paragraph{Rotational and translational motion of active polar droplets.}
We will now consider a case with weaker anchoring to \textcolor{black}{the} droplet interface. Under these conditions, the equilibrium configuration in absence of activity does not exhibit anymore cylindrical symmetry, regardless of the kind of anchoring considered. The polarization pattern for tangential anchoring (panel (e) of Fig. \ref{fig:5})  exhibits a spindle-like shape with no topological defect in the droplet bulk. This is possible because tangential anchoring is enforced everywhere at the droplet boundary but at the two vertices of the spindle (see green circles in panel (e)). This occurs because of energetic reasons as it is more favorable not to respect the anchoring rather than producing a topological defect.
The equilibrium configuration for homeotropic anchoring can be obtained from that with tangential anchoring one by applying a global rotation of $\pi/2$ to the latter.
We begin by considering the case with extensile activity and homeotropic anchoring, as in the previous section.
Fig. \ref{fig:5}(a)-(d) represents two different dynamical regimes which can be achieved by only varying the rate of active injection $\zeta>0$.
At low activity the droplet is in a quiescent state where it slightly elongates in the mean direction of polarization (panel (a)). The flow generated by the distortion arranges in 4 vortices with null total angular momentum (panel (b)). By increasing activity the droplet undergoes a transition towards a motile state. The droplet deforms, attaining the shape of a tilde (see panel (c)) and the flow structure develops a stationary vortex (see panel (d)). Interestingly enough, the transition from the quiescent to the rotating regime is mediated by the appearance of an oscillating regime where the droplet periodically switches from clockwise to anticlockwise rotation --a behavior reminiscent of those observed in bulk active fluids close to a non-equilibrium phase transition~\cite{denniston2001}.  
Both transitions, between quiescent and oscillating state first, and  between oscillations and steady rotations, are characterized by  singularities in the behaviors of rotational velocity~\cite{fialho2017}.

We now focus on the case of an extensile droplet with tangential anchoring. Switching the anchoring features of the liquid crystal has a drastic effect on the droplet dynamics. Indeed, for large enough activity the singular points at droplet interface rearrange to maximize the bending in the bulk by closing in at the rear (see Fig. \ref{fig:5} (f)). Such a configuration is not axisymmetric and the liquid crystal deformation produces a net flow in the direction normal to the bending. As a result, the droplet sets into a stationary translational motion, whose speed is found to be directly proportional to the intensity of the activity $\zeta$~\cite{fialho2017}.

What happens if contractile activity is considered? In this case the liquid crystal is unstable to splay deformations so that the liquid crystal patterns obtained with tangential and homeotropic anchoring are basically reversed with respect to the extensile case. 
Fig. \ref{fig:5} provides a schematic representation of the anchoring-driven dynamics observed in $2D$ active droplets. 
For instance, in contractile droplets tangential anchoring
stabilises two regions of bend distortions, generating two forces which counterbalance each other, so that no translation can occur.
However, if these are not collinear they give rise to an active torque which eventually fuels the rotation of the droplet~\cite{fialho2017} (see panel (l) of Fig. \ref{fig:5}).
Conversely, when homeotropic anchoring is considered, the splay instability leads to only one region of coherent deformation generating a non vanishing force (but no torque) so that the droplet translates without rotating~\cite{fialho2017}. 
\textcolor{black}{The same is true for contractile droplets with homeotropic anchoring~\cite{Giomi2014} with nematic rather than polar order (see panels (g)-(h) of Fig. \ref{fig:5}).}
It worths noticing that in this  case, under the right conditions for the elastic constant and surface tension of the droplet, Giomi and De Simone~\cite{Giomi2014} have shown  that the droplet may spontaneously split (see panel (i) of Fig. \ref{fig:5}) for large enough contractile activity --a behavior which is relevant to understand cell division.

\begin{figure}[t!]
\center
\includegraphics[width=0.9\textwidth]{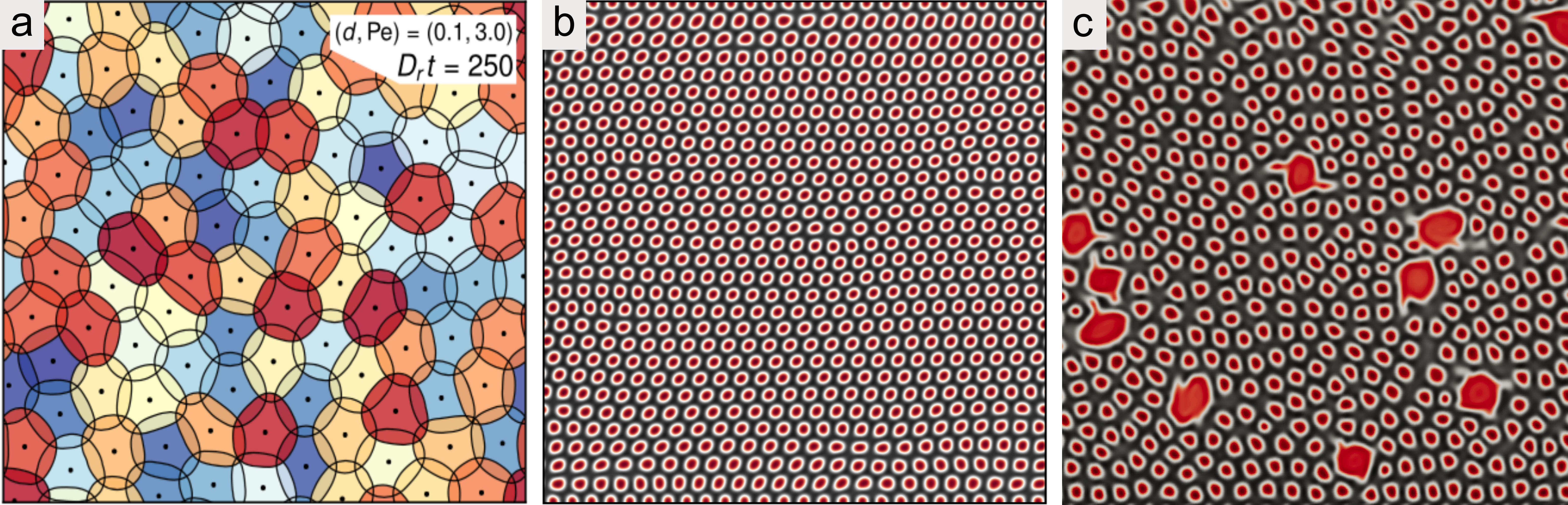}
\caption{ \textbf{Systems of multiple droplets.} Panel~(a) shows the configuration of a monolayer of active droplets obtained through multiphase simulations(courtesy of Prof. Davide Marenduzzo and Michael Chiang). Black lines and dots define the contour of each cells and their centres of mass respectively. The color code refers to the coordination number of each cell. At small and vanishing activity, droplets arrange into a stationary hexatic phase (not shown), while at larger activity the tissue undergoes a \emph{melting} transition and droplets start to exchange neighbors.  Panels~(b) and~(c) show two configurations at small and large activity respectively, obtained from simulations of surfactant models. In the quiescent state of panel~(b) an array of droplets arrange into a hexagonal lattice with some dislocations scattered throughout the system. At larger activity, droplets begin to merge into larger domains whose size may be large enough to produce the bending instability discussed in Section~\ref{sec:3.4}. In this case the larger droplet rotate under the fuelling effect of active injection, a behavior which closely reminds the dynamic of bacterial suspension on a substrate~\cite{basaran2020}.}
\label{fig:6}       
\end{figure}

\textcolor{black}{
\paragraph{Tissues and Active emulsions.}
A promising field of investigation, both from an experimental and numerical point of view, concerns the case of active emulsions, namely systems where multiple active droplets coexist and interact with each other (Fig.~\ref{fig:6}). These systems may potentially have important applications ranging from designing devices for drug delivery to novel biomimetic smart materials able to resist to intense deformations~\cite{sanchez2012,herminghaus2014}.  
Here we want to highlight two possibile routes to adapt the active gel theory to take into account system with multiple droplets. First, one may use a multiphase approach, where each droplet is described by a different concentration field. Another possibility is to consider surfactant models to arrest phase separations and stabilize an emulsion of liquid crystalline active droplets in an isotropic background\footnote{For instance this can be done by using negative values for $k_\phi$ in Eq.~\eqref{eqn:phi4}.}. The first approach allows to achieve arbitrary packing fractions and, for this reason, it has been used in particular to study the dynamics of confluent tissues. Surfactant models are more suitable to study self-separating systems  instead (\emph{i.e.} bacterial colonies) and ensure a better computational performance, enabling to study the long term dynamics~\cite{bone2017,negro2018} and rheological properties of such systems~\cite{negro2019,carenza2020}.
}

\paragraph{Self-propelled droplets without liquid crystalline order.}
In previous Sections we considered a series of realizations of active fluids based on the active gel theory, where the elasticity of the internal filamentous structure of the active constituents was taken into account. We briefly mention here that activity can be introduced in a continuum model even in absence of liquid crystalline order.

The simplest realization of such a model consists of a single diffusive, conserved order parameter $\phi(\mathbf{r},t)$. In the passive limit the evolution of such a field is given by the \emph{model H} (in the classification of Hohenberg and Halperin~\cite{hohenberg1977}) consisting of the system of partial differential equations Eq.~\eqref{eqn:Cahn_Hilliard}, Eq.~\eqref{eqn:nav} subject to the incompressibility condition of Eq.~\eqref{cont}. In such a scalar model, the stress $\sigma$ is the sum of the viscous contribution in Eq.~\eqref{eqn:viscous} and the interfacial term:
\begin{equation}
\sigma_{\alpha \beta} = - \kappa_\phi \left( \partial_\alpha \phi \partial_\beta \phi -\dfrac{1}{d} (\nabla \phi)^2 \delta_{\alpha \beta} \right),
\end{equation}
where $d$ represents the dimensionality of the system. If $\kappa_\phi = k_\phi$ appearing in the free energy reported in Eq.~\eqref{eqn:phi4} the stress $\sigma$ coincides with the equilibrium interfacial stress.
However, one may allow $\kappa_\phi$ to attain general values different from $k_\phi$, so that $\kappa_\phi = k_\phi + \zeta_\phi$, where $\zeta_\phi$ plays the role of an \emph{acitve parameter} for the scalar theory.
This model, known as \emph{active model H}, was first proposed by Tiribocchi \emph{et al.}~\cite{tiribocchi2015} in 2015 and it has proved successful to model the cortical activity of cells in tissues.
More elaborate models including two scalar fields or variable density were also developed to study the motility features of active droplets in absence of internal liquid crystalline order~\cite{Negro2019_EPL,PhysRevResearch.2.032024}.

\section{Topology of Liquid Crystal Droplets in 3D}
\label{sec:4}
In the second part of this Chapter we will focus on full 3D realizations of active droplets and shells with liquid crystalline order. Some results presented in the previous sections for bidimensional systems also apply in 3D --\emph{i.e.} LC crystal deformations fuel internally driven motion-- but the increased degrees of freedom due to the presence of an extra dimension has a drastic effect on the topological rules which confined liquid crystals have to match and consequently, on the resulting dynamics of active droplets. 
Indeed, in the next Section we will present a number of situations where topology plays a leading role in selecting the dynamical behavior of the system. However, before getting involved into the analysis of active droplets, we will here provide a concise overview of some topological facts relevant for the proceeding of our discussion.

In Section~\ref{sec:activegeltheory} we already discussed how liquid crystals can develop smooth deformations in the form of bending, splay or twist (Fig.~\ref{fig:2}). However, when distortions strenghten, the elastic pattern may give rise to singular
regions --namely topological defects-- where it is not anymore possible to define a definite preferential direction of alignment. The notion of topological defect goes far beyond the theory of liquid crystals as it is a
common feature of most system with broken isotropic symmetry~\cite{Teo2017,Kosterlitz2017}. \textcolor{black}{For instance, breaking of translational order in crystalline solids gives rise to \emph{dislocations}, while in liquid crystals \emph{disclinations} are regions which break orientational order, in the same way as \emph{vortices} in superfluid helium.}
The structure of topological defects strongly depends on the symmetries of the system and its dimensionality and they may either develop in the form of point-like regions or lines embedded in a higher-dimensional space.

\label{sec:3.3}
\paragraph{Disclinations in two dimensions.}

\begin{figure}[t!]
\center
\includegraphics[width=0.74\textwidth]{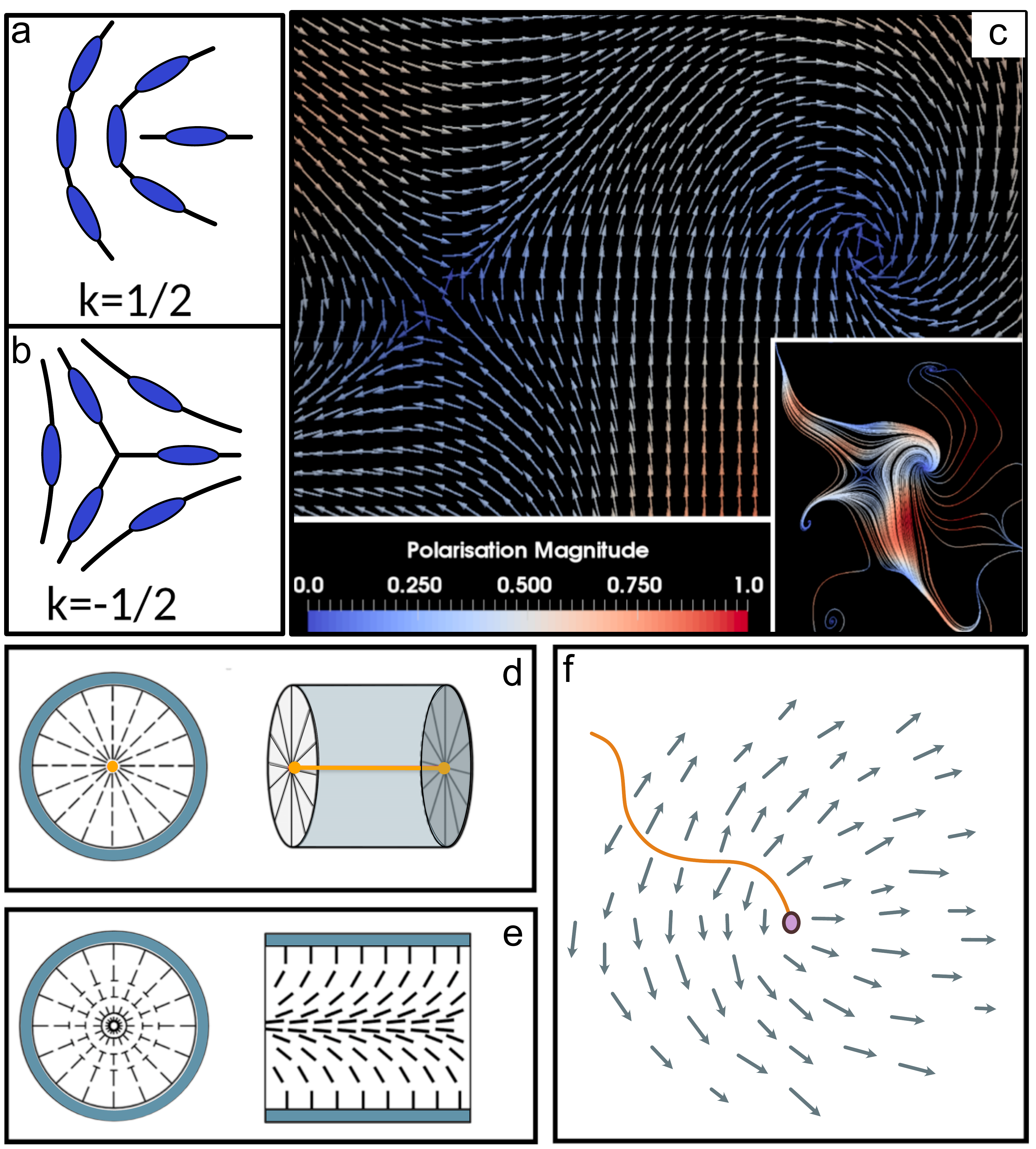}
\caption{ \textbf{Disclinations in 2D and stability of disclination lines in 3D.} (a)-(b) Cartoons of two semi-integer defects of charge $+1/2$ and $-1/2$ respectively, in the bidimensional space. Semi-integer defects are allowed only in systems with nematic symmetry for topological reasons. Systems with polar symmetry can develop defects with charge $\pm 1$ (and multiples). These are shown in panel~(c) where red arrows correspond to ordered regions, while blue arrows are found in proximity of topological defects signalling that the polarization magnitude is approximately null and order is locally lost. A vortex with $+1$ charge and an anti-vortex with $-1$ charge are formed respectively on the right and left of the panel. Inset shows the streamlines of the vector field in proximity of the
two defects. (d) Schematic representation of
a $k = 1$ disclination line (in red), characterized by an aster configuration in the normal planes, as shown in the right of the panel. However, disclination lines with integer winding number are not topologically stable, since the order parameter can always escape in the third dimension as shown in (e). (f) Volterra construction for a $+1/2$ topological defect in 2D, characterized by a branch cut (red line) in correspondance of those points where the orientation of the vector field changes abruptly. This configuration is also analogous to the one developed in 3D on planes normal to a disclination line with $k=1/2$. Therefore, a disclination line is defined as the continuation (in the direction normal to the plane in the figure) of the edge of the branch cut.
}
\label{fig:7}       
\end{figure}

A disclination can be characterized by looking
at the configuration of the order parameter far from its
\emph{core} --\emph{i.e.} the region where the notion of alignment is lost. This can be done by computing $k$ the \emph{winding number} or \emph{topological charge} which is a measure of the strength
of the topological defect and is defined as the number of
times that the order parameter turns of an angle of $2\pi$
while moving along a close contour surrounding the defect
core~\cite{chaikin1995}. 
Fig.~\ref{fig:7}(a)-(c) shows a collection of topological defect in the plane
with different charge.
Importantly, the energy of such configurations can be expressed in the single constant approximation as 
\begin{equation}
f_{def}=\pi K k^2 \ln \left( \dfrac{R}{r_c} \right),
\label{eqn:energy_defect}
\end{equation} 
where $r_c$ is a short-distance cutoff which represents the defect core, while $R$ is a large-scale cutoff of the order of the distance to the boundary or other defects. The quadratic dependence on the defect charge plays an important role on the physical stability of defect configurations since high-charged disclinations pay a greater energy cost than low-charged defects.

The relevance of disclinations 	in the study of LC lies in their connection with the \emph{global} topological properties of the system. Indeed, while smooth distortions, as the ones shown in Fig.~\ref{fig:2}, can be returned to the ground state through a series of infinitesimal deformations of the order parameter, disclinations cannot be cancelled with such a procedure, and some form of \emph{surgery} is necessary, thus requiring a discontinuous change in the direction of alignment (see for instance panel~(f) of~Fig.~\ref{fig:7}). However, topological defects in $2d$ can be removed through a process of \emph{annihilation} between defects of opposite charge. Indeed, one may consider a region populated by two oppositely charged defects--as the ones shown in Fig.~\ref{fig:7}(c). By taking a contour which surrounds both defect cores, it is immediately found that the topological charge enclosed in the contour is null, thus signaling that such configuration can be continuously deformed into a perfectly aligned state.  
Generally, there is no \emph{topological} rule which prevents either the formation or the elimination of disclinations, as long as the total topological charge of the system is \emph{conserved}. This implies that the dynamics of annihilation/enucleation of topological defects is a \emph{pair process}. Analogously, a single topological defect of charge $k$ may also split into two topological defects with charge $k_1,k_2$ provided that $k=k_1+k_2$.
For instance a $+1$ vortex may split into two $+1/2$ defects, thus allowing for a reduction of the total energy of the system, according to 
Eq.~\eqref{eqn:energy_defect}. 

However, this energetically favoured decaying process is \emph{topologically} allowed only for liquid crystals with nematic symmetry. Conversely, polar LC can only develop defects with integer charge, while semi-integer configurations are forbidden due to the lack of inversion symmetry. This important remark will play an fundamental role in the next paragraphs where we will discuss the topological properties of disclination lines in higher dimensionality.


\paragraph{On the concept of topological stability.}
Defects are said to be topologically stable if a non uniform configuration of the order parameter cannot be reduced to a uniform state by a continuous transformation.
A general criterion to establish whether a defect is topologically stable or not, is to look at the dimension $n$ of the order parameter. In a $d$-dimensional space, the condition that all the $n$ components of the order parameter must vanish at the defect core defines a \emph{hyper-space} of dimension $d-n$. Hence defects exist if $n\leq d$. In the fully two dimensional case ($d=2$) of the previous Section the order parameter are free to orient in the plane ($n=2$), and the defects allowed are points (or vortices). However,  point defects can be unstable in quasi-$2d$ systems, \emph{i.e.} when the order parameter fully lives in the three-dimensional space, as in such case one would have $n>d$: indeed the director field in proximity of a vortex is always capable to escape out of the plane aligning with itself, thus removing the defect (see for instance panel~(d) and~(e) of Fig.~\ref{fig:7}). 
However, the condition $n\leq d$ defines a necessary but not sufficient condition for the stability of a defect configuration, as it will become clear in the next Section.

\paragraph{Hedgehogs and disclinations in the three-dimensional space.}

\begin{figure}[t]
\center
\resizebox{.8\textwidth}{!}{%
  \includegraphics{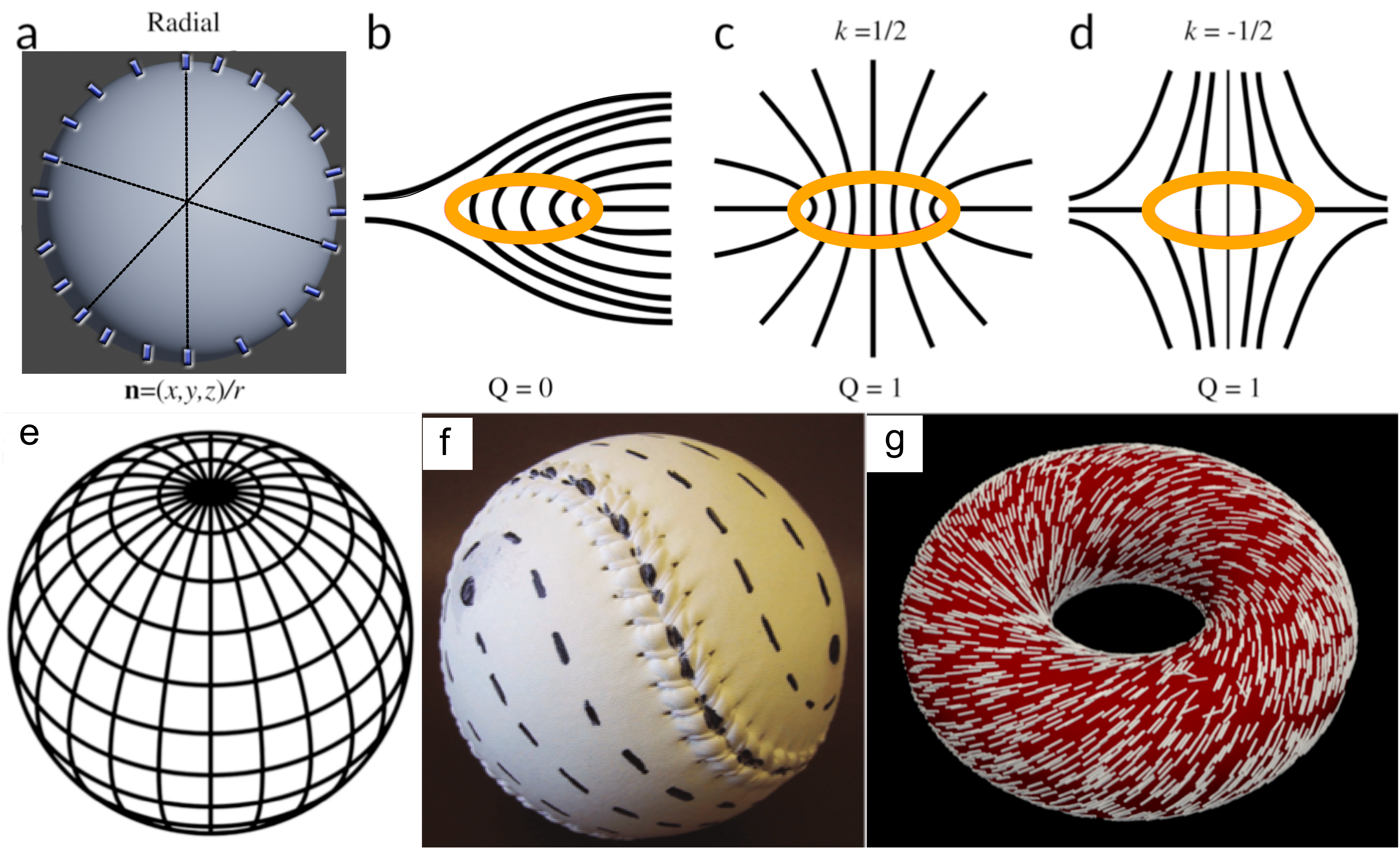}
}
\caption{ \textbf{Hedgehogs, disclination loops and topological defects in non-Euclidean topologies.} Panel (a) shows the typical radial configuration of a hedgehogs. Panel (b) shows a disclination loop with charge Q = 0 (red ring). In black the director streamlines in the plane normal to the disclination line are shown. Panels (c) and (d) show two $+1$ disclination loops respectively with winding number $k =1/2$ and $k = 1/2$ and hedgehog topological charge $Q=1$. These two configurations are topologically equivalent as they can be obtained by continuously deforming one into the other. Panel~(e) shows how the poles on the globe are singular points of  meridian and parallels with positive unitary charge. Two defects are thus necessary to satisfy the condition imposed by the Poincaré-Brouwer theorem in Eq.~\eqref{eqn:poincare}, since the genus of the sphere is
null. Conversely, four $+1/2$ defects disposed on the vertices of a tetrahedron are necessary to draw the typical pattern of a baseball ball, shown in panel~(f). (g) Nematic LC on a torus arranges in a defect-free configuration, being the Euler characteristic null for this geometry. Picture has been obtained through lattice Boltzmann numerical simulations by the authors. Panels (a-d) adapted from~\cite{Lubensky1998}.}
\label{fig:8}       
\end{figure}

In three-dimensional systems ($d=3$) LC may  either have  point (\emph{hedgehogs}) or line defects. \textcolor{black}{The former are point defects characterized by the fact that the far-field configuration of the director field visits every point on the unit sphere (in a similar fashion as the electric field generated by a point-like charge). Their topological charge $Q$ (often addressed as \emph{hedgehog charge} to distinguish it from the winding number) is calculated by computing how many time\textcolor{black}{s} the director field on a closed surface enclosing the defect wraps the unit sphere.}
Instead, a line defect is characterized by a core around which the director winds, giving rise to patterns analogous to those shown in Fig.~\ref{fig:7} on planes normal to the disclination line. However, in $3d$, defects with $k=\pm1/2$ and  $k=\pm 1$ have a radically different nature. Indeed, only the former are topologically protected.
Intuitively, it is reasonable that the $k=1$ disclination line configuration of Fig.~\ref{fig:7}(d) may be continuously deformed into the one shown in panel~(e) by letting the director field escape in the third dimension.

Yet, for  disclination lines with semi-integer winding number the situation is  different. Indeed if any semi-integer disclination exists on the boundary of a LC sample, this must be connected to a disclination line which pierces the bulk. To better explain this point, we consider the (bidimensional) $+1/2$ disclination in Fig.~\ref{fig:7}(f). Here, the director field $\mathbf{n}$ features a branch cut, denoted with a red line, where it abruptly changes \textcolor{black}{by} an angle $\pi$. 
This phase change has no physical relevance, since nematic symmetry prescribes that the two orientations $\pm \mathbf{n}$ are actually equivalent. However, the existence of a branch cut also prevents from the possibility of eliminating a semi-integer disclination by escaping in the third dimension~\cite{Meyer1973}. This is because it is not possible to eliminate the branch cut by means of any series of continuous transformations. For instance, we may try to pull the director out of the plane of the figure through a positive rotation around the axis defined by the centre of the defect and the position of the director. At the end, we would have a configuration where the vectors on the right of the branch cut point downwards and those on the left point upwards.
Therefore, if we now suppose that the configuration of panel~(f) is actually on the surface of a three-dimensional system, it becomes obvious that the branch cut must penetrate the bulk thus generating a smooth surface whose edge defines the $+1/2$ disclination line (which is the only physical region of discontinuity).
Therefore, there is no global transformation which may continuosly eliminate the branch cut, as discussed in the previous Section.
Conversely, in the case depicted in Fig.~\ref{fig:7}(d) for a $+1$ disclination line, no branch cut occurs, and it is therefore possible to eliminate the singular line through a series of continuous transformations. It follows that integer (point-like) surface defects exist as isolated structures and are not connected to any disclination in the bulk.


Such analysis goes beyond the case discussed here for $+1/2$ and $+1$ defects, since any semi-integer disclination requires a branch cut while any integer disclination does not. Hence, the previous discussion suggests that disclination lines are always associated with semi-integer defect patterns, while hedgehogs can only develop integer defect patterns (see for instance the radial pattern in Fig.~\ref{fig:8}(a)). This also has \textcolor{black}{an} important effect on the topological defect structure expected in nematic and polar LC. Indeed, while both point and line disclinations can develop in the former, only point defects can develop in polar LC since their vector nature prevent the formation of disclinations with semi-integer winding number.

		
\paragraph{Disclination loops.}

Another fundamental difference from the bidimensional case is that, while the two oppositely charged semi-integer defects in panels (a) and (b) of Fig.~\ref{fig:7} are topologically different, in three dimensions they are homotopic --\emph{i.e.} they represent the same topological object, since the third dimension allows for a mapping of the two configurations through a continuous transformation.

This has important implications in the characterization of closed disclinations, namely disclination loops.
A schematic representation of three possible loop configurations is shown in Fig.~\ref{fig:8}(b-d).
The  loop disclination in panel~(b) is characterized by a continuous deformation from a disclination pattern with winding number $k=+1/2$ into one with $k=-1/2$. However, far enough from the loop, the director field is broadly uniform and the far-field configuration is not influenced by the presence of the disclination.
Conversely, the two homotopic loops in panel~(c) and~(d) lead to director configurations in the far field which visit every point of the projective unit sphere. Thus, they are topologically equivalent to the radial hedgehog in panel~(a) and for this reason they can be labeled with a hedgehog topological charge $Q=1$~\cite{Mermin1979}.

		
\subsection{Topological defects in Non-Euclidean Manifolds}
\label{sec:noneuclidean}
In the previous Section we provided an overview on the main topological 
features of disclinations in the bulk of a nematic LC sample.
However, in many practical situations LC are confined in finite geometries where topological properties on the surface may importantly alter the arrangement of the LC in the whole sample.

More specifically, it is interesting to consider the case of a LC confined 
at the interface of a closed $2d$ manifold embedded in the three-dimensional space. Under these circumstances, the arrangement of the director field is not determined only by thermodynamic forces,  since the topological properties of the LC must match the global topological properties of the curved manifold~\cite{Kamien2002,rassegnadifettiliquidcrystalsgocce}.
The most striking example of such mathematical concept is provided by the impossibility of combing an \emph{hairy ball}: no matter what you try, eventually you would always end up with a configuration characterized by some \emph{hairy whorl} --hence a topological defect.
This is for instance the case of meridians on the Earth where the poles are
$+1$ point defects on a sphere surface with total topological charge $+2$ (see for instance panel~(e) in Fig.~\ref{fig:8}).
All of these are realizations of the famous Poincar{\'e}-Brouwer theorem~\cite{Eisenberg1979} which states that the total topological defect charge must be equal to the Euler characteristic $\chi$ of the closed orientable manifold
\begin{equation}
\sum_i^{N} k_i = \chi,
\label{eqn:poincare}
\end{equation}
where here $N$ denotes the number of disclinations.
Moreover, $\chi$ is connected to the global properties of the Gaussian curvature, hence to the genus $g$ of the manifold (the number of handles) through the following relation\footnote{Notice that this relation is only valid for closed orientable manifolds. In the case of other topologies, as for instance the infinite plane or a disk, the Euler characteristic can be computed by finding a polygonization of the surface.}~\cite{Kamien2002}
$$\chi=2(1-g).$$
This result is stating, for instance, that the director
field on the surface of a sphere ($g=0$) must have one $+2$ defect (or two $+1$ defects, as well as four $+1/2$ defects). On a torus ($g=1$), however, no defects
are necessary. 
A gallery of geometries with different Euler characteristic is provided in Fig.~\ref{fig:8}(e-g).

\begin{figure}[t!]
\center
\resizebox{.82\columnwidth}{!}{%
  \includegraphics{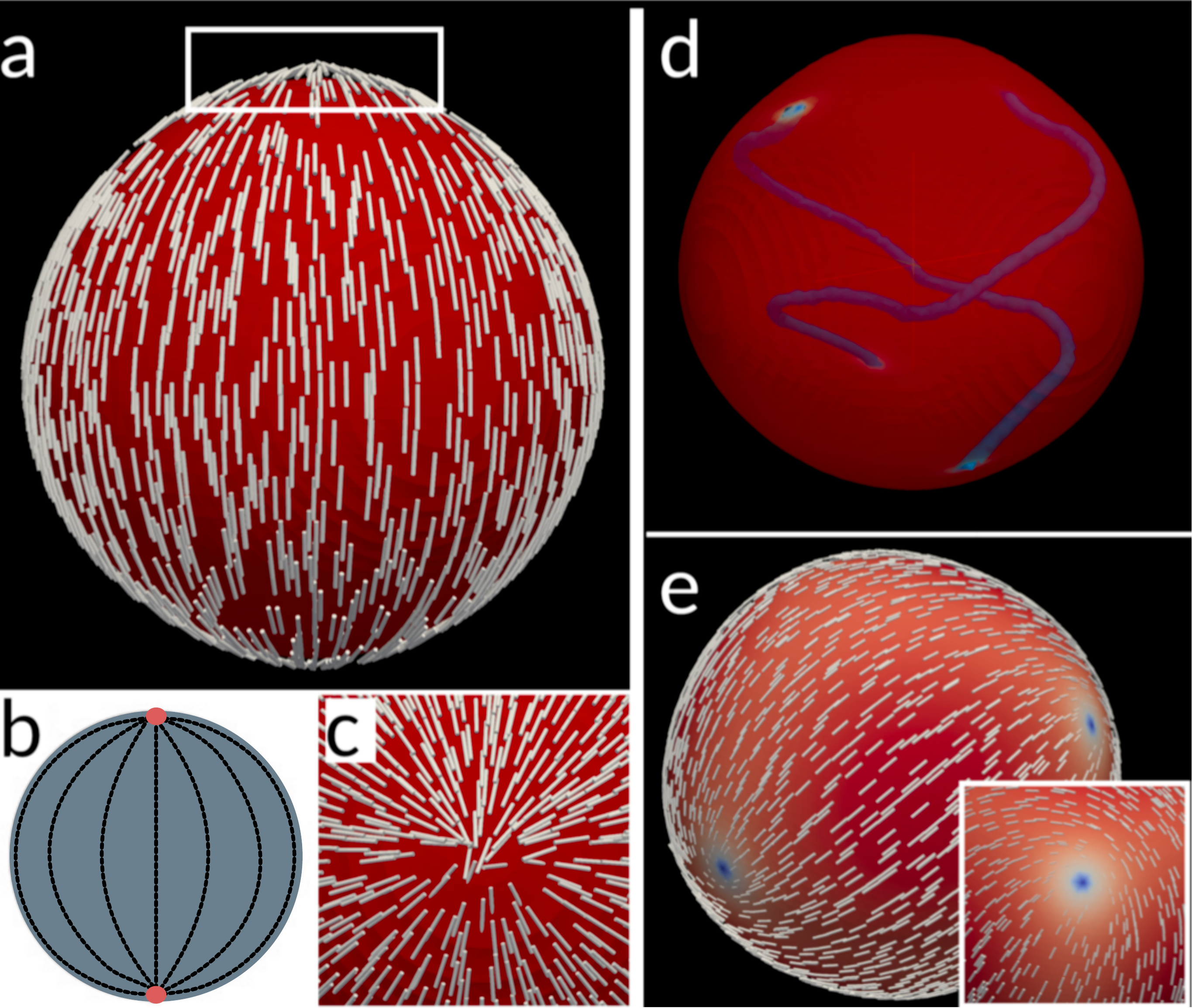}
}
\caption{ \textbf{Droplets and shells of liquid crystal.} (a) Bipolar configuration in a LC droplet with tangential anchoring at the interface. Two antipodal boojums with +1 charge are formed at the droplet poles. (b) Schmeatic representation of the internal nematic arrangement in the bipolar configuration. (c) Zoom on the director arrangement in correspondace of the boojum highlighted with a white box in panel (a). Panel (d) shows a non-equilibrium configuration where four $+1/2$ defects (blue spots on the surface) are formed at the droplet interface. In this case disclination lines (navy blue lines) pierce the interior of the droplet joining topological defects in pairs. (e) Equilibrium tennis-ball configuration in a thin shell of LC with four $+1/2$ defects equally spaced on the droplet surface on the vertices of a tetrahedron. Inset shows the arrangement of the director field in proximity of a defect core.
} 
\label{fig:9}       
\end{figure}

\paragraph{Droplets and shell of liquid crystals with tangential anchoring.}
\label{sec:model_simulation}

Although the topological charge is fixed, liquid crystal droplets with tangential anchoring of the director at the interface may yet develop various equilibrium configurations, depending on the relative intensity of the elastic constant in the Frank free-energy in Eq.~\eqref{eqn:Frank_nem}.

A possible equilibrium state is the bipolar structure, shown in panel (a) of Fig.~\ref{fig:9}, where two antipodal $+1$ defects --namely \emph{boojums}~\cite{mermin1990}-- are found at the droplet poles where the director attains an aster-like pattern (see for instance panel~(c)).
Hence, such configuration is characterized by splay distortions in proximity of the defect core and bending everywehere else, as shown by the cross-section in panel~(b).

Under the action of some kind of forcing, each of the two boojums may split into two defects with $+1/2$ charge. As discussed above, defects with semi-integer charge cannot be isolated and must be connected to a disclination line. This is the case of the (non-equilibrium) configuration in panel~(d)~\cite{Carenza22065} where two disclinations pierce the droplet joining the surface defects in pairs. These introduce further deformations in the droplet bulk which result in a greater energy cost if compared to the bipolar state. Hence, in absence of forcing, the four defects would rearrange and merge in pairs to reconstruct the bipolar configuration, leading to the vanishing of the two disclinations in the bulk.
However, if a shell of LC is considered, rather than a droplet, one finds that the preferential equilibrium state is given by the baseball ball configuration in Fig.~\ref{fig:9}(e), where four $+1/2$ defects are placed on the vertices of a regular tetrahedron. In this case, no disclination line is formed, since the liquid crystal is confined at the droplet interface in a bidimensional fashion and the tetrahedrical configuration is the one which best minimizes splay deformations on the sphere~\cite{guillamat2018}.

\section{Active droplets and shells in 3D}
\label{sec:5}
In this Section we will present a series of results concerning the motility properties of active liquid crystals in confined geometries. Following the chronological order, we will start from the case of active droplets and we will conclude with acive shells.

\begin{figure}[ht!]
\center
\includegraphics[width=0.65\textwidth]{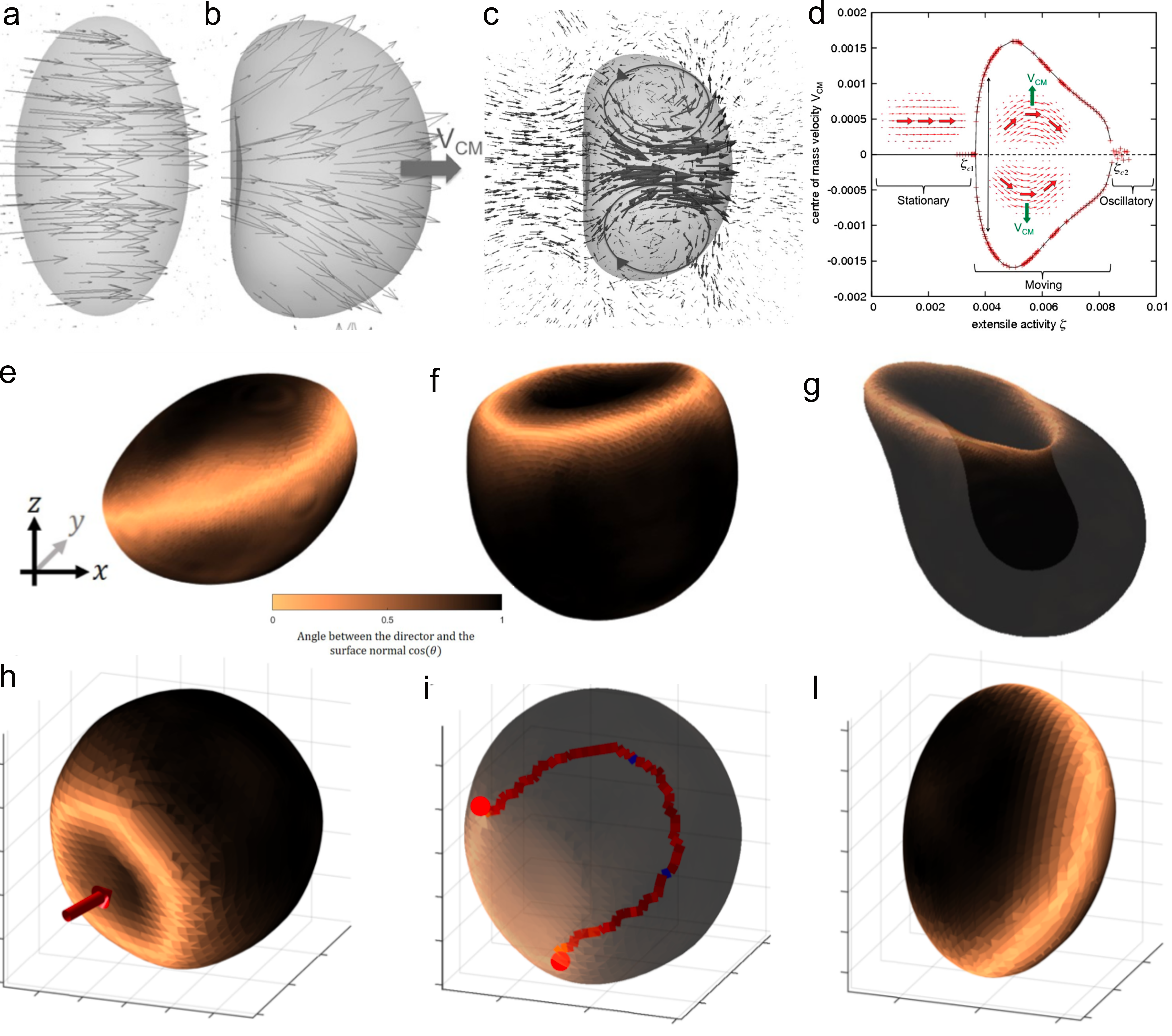}
\caption{ \textbf{Active droplets in 3D.} \textit{\textbf{(a)-(d)} Active polar droplets in 3D.} In panels (a) and (b) thin grey arrows show the polarization configuration. For small values of contractile activity (panel (a)) the polar droplet contracts in the direction of the mean polarization. For stronger forcing the droplet is no more symmetric (panel (b)) and starts to migrate with a flow field shown in panel (c) (thick arrows), featuring two vortices rotating in opposite directions whose effect sums up at the droplet centre sustaining the translational motion of the droplet. (d) Center of mass velocity of the droplet vs activity. \textit{\textbf{(e)-(l)} Active nematic droplets.} Colour code indicating the local surface alignment of the director field. Panels (e)-(g) show the time evolution of the invagination process descirbed in the text. (e) The ring of in-plane alignment at the equator (light region) causes nematic bend deformations in the bulk which are pushed outwards by contractile activity, thus deforming the droplet to an oblate shape. (f) While the ring of in-plane alignment moves towards one of the poles, the bend deformations in the bulk point perpendicular to the surface which causes the active forces to develop a component pointing towards the pole thus creating a dent in the centre of the ring. (g) Contractile activity  causes  splay deformations to push inwards, thus increasing the depth of the dent. Panels (h)-(l) show a droplet performing a run-and-tumble motion. This happens if the interface is too stiff for active forces to cause complete droplet invagination.  In this case the bend stripe at the equator shrinks forming a dip  at the centre of the bend-ring where a splay deformation in the bulk pushes the droplet forward (h). The elastic deformations relax via the nucleation of a  half- loop (i). The self-propulsive disclination line moves through the droplet and eventually reaches the interface. Thereby the bend-ring is re-oriented along a random direction and the process repeats (l).  \textcolor{black}{Panels (a)-(d) are adapted from~\cite{Tjhung2012} with permission from D. Marenduzzo. Panels (e)-(l) are adapted from~\cite{ruske2021}with permission from J. Yeomans.}
} 
\label{fig:10}       
\end{figure}

The first paper where the case of a droplet in the full 3D space was considered appeared in 2012. Here, Tjhung \emph{et al.}~\cite{Tjhung2012} numerically studied an active polar droplet in absence of anchoring\footnote{A similar model was also used to study numerically the crawling motion of cells on a substrate~\cite{tjhung2015}.}, showing that self-motility results from the spontaneous symmetry breaking of polarity inversion symmetry. In particular, they showed that stable translational states develop in presence of contractile stresses. Indeed, at small activity the droplet contracts in the direction of the mean polarization and slightly elongates in the normal direction (panel (a) of Fig.~\ref{fig:10}) with an internal quadrupolar flow which counteracts surface tension and stabilises the elongated shape, without net motion. At large enough, yet moderate, contractile activity this state becomes unstable to splay deformation, as in its 2D counterpart which we analyzed in the previous sections. The resulting state is either characterized by positive splay if the polarization fans outwards as shown in Fig.~\ref{fig:10}(b) or negative splay if the polarization direction is reversed. As the spontaneous symmetry breaking occurs, the droplet starts to migrate in the direction of the active force, which is set by the sign of $\nabla \cdot \mathbf{P}$. The symmetry breaking also affects the flow field which now only features two vortices rotating in opposite directions whose effect sums up at the droplet centre sustaining the translational motion of the droplet (see panel (c) of  Fig.~\ref{fig:10}). Such spontaneous symmetry breaking can be seen from a dynamical perspective as a continuous non-equilibrium phase transition consistent with a supercritical Hopf bifurcation, as suggested by Fig.~\ref{fig:10}(d) where the velocity of the centre of mass is plotted against the activity parameter $\zeta$.

How does this picture change if a liquid crystal with nematic (rather than polar) symmetry is considered? The question has been recently addressed by L.J. Ruske and J.M. Yeomans in~\cite{ruske2021} where they studied the case of active nematic droplets in \textcolor{black}{the} absence of any thermodynamic anchoring. In particular, they were interested in the dynamics of deformable droplets, \emph{i.e.} when the activity is large enough to counteract the droplet surface tension leading to shape change. While the underlying mechanism leading to the active instability is the same as the one unveiled by Thjung \emph{et al.} for polar droplets, yet the topological features of nematics lead to a collection of various interesting behaviors.


This variety was explained by considering that activity leads to preferential director alignment at the droplet interface, even in absence of thermodynamic anchoring. This phenomenon, already known in 2D realizations of active droplets, has a drastic  influence in 3D. In particular, they found that both extensile and contractile activity leads to an effective alignment at the interface, also called active anchoring,  which is tangential for the former and homeotropic for the latter~\cite{ruske2021}. 
Focussing on the case of contractile droplets, Ruske and Yeomans observed that activity-induced normal anchoring would eventually result into a ring of in-plane surface alignment (see panel (e) of Fig.~\ref{fig:10}), called \emph{bend-ring}, topologically equivalent to a +1 hedgehog or disclination loop but, energy-wise, less expensive.
Contractile activity pushes the ring outwards, deforming the droplet into an oblate ellipsoid. At small activity the shape change is stabilized by surface tension, however at larger activity the equatorial configuration of the bend-ring is unstable and the latter migrates towards one of the poles (see panel (f) and (g) of Fig.~\ref{fig:10}) --a process which closely resambles Tjhung's spontanous symmetry breaking. According to the relative importance of active stresses and surface tension this instability may result to either the \emph{invagination} of the droplet (initiated by the mutual effect of bending in proximity of the bend-ring and splay in the rest of the droplet) or \emph{run and tumble} if the interface is too stiff to initiate the invagination process (panels (h)-(l) of Fig.~\ref{fig:10}). In this latter case, however, the bend-ring relaxes giving rise to a disclination line which autonomously moves inside the droplet (panel (i) of Fig.~\ref{fig:10}) until it reaches the opposite emisphere where it re-orients the bend-ring (panel (l)) and the dynamics repeats in a random direction.
For extensile droplets tethering formation was observed.

\subsection{Active nematic droplets with tangential anchoring: spontaneous rotation}
\begin{figure}[t!]
\center
\includegraphics[width=0.8\textwidth]{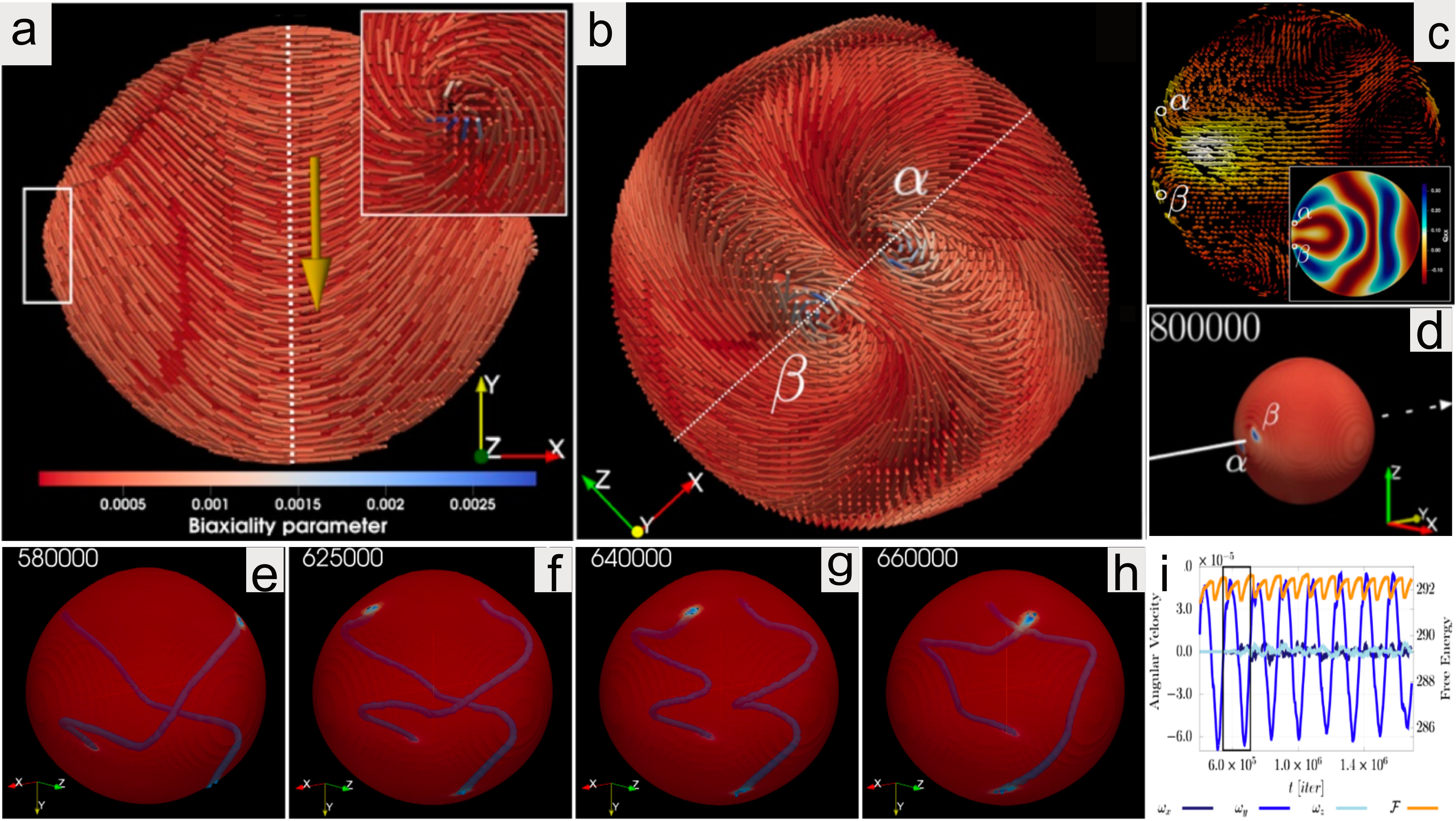}
\caption{ \textbf{Anchoring and chirality effects on active droplets motility} \textit{\textbf{(a)} Extensile Nematic droplet with tangential anchoring.} Panel (a) shows the configuration of the director field  on the droplet surface. Two stationary $+1$ boojums are formed at antipodal points in the $x$ direction. Inset of panel (a) shows the director field in proximity of the boojum framed with a white box. Bending deformations occur transversally to the long axes of the droplet and generate an active force in the direction of the yellow arrow, thus powering rotational motion in the $yz$ plane.  \textit{\textbf{(b)-(d)} Screw-like propulsion in a chiral extensile droplet}. Panel (b) shows the configuration of the director field on the droplet surface.  Two +1 defects are labeled with Greek letters $\alpha$ and $\beta$.  The screw-like rotational motion generates a strong velocity field in the interior of the droplet in proximity of the 2 defects. The velocity field is plotted in panel (c) on a plane transversal to the plane of rotation of the 2 defects (panel (d)) represented  by the dashed line in panel (b).  Inset of panel (c) shows the contour plot on the same plane of the $Q_{xx}$ component of the $Q$ tensor, exhibiting an arrangement similar to the radial spherical structure~\cite{rassegnadifettiliquidcrystalsgocce}. \textit{\textbf{(e)-(i)} Disclination dance of a cholesteric droplet with inward active torque dipoles.} Panels (e)-(h) show snapshots of the droplet and the disclination lines. The four $+1/2$  defects (represented by blue dots on the droplet surface) rotate in pairs in opposite directions (top defects rotate anticlockwise, while bottom defects rotate oppositely). As the defects rotate, the $2$ disclination lines first create a link (f); then, they recombine (g) and finally, relax into a configuration close to the initial one (h) but switched. (i) The angular velocity over time, oscillates from positive to negative values. Here, the time evolution of the free energy (orange curve) shows that the free energy oscillates with a double frequency. Figure adapted from~\cite{Carenza22065}.} 
\label{fig:11}       
\end{figure}

By including tangential thermodynamic anchoring the dynamics of active nematic droplets changes drastically.
Carenza \emph{et al.} showed that even in \textcolor{black}{the} presence \textcolor{black}{of} anchoring, activity can be used to select different regimes~\cite{Carenza22065}. At low values of $\zeta$  the droplet is static, and the director field attains a boojum-like pattern (see panels (a-c) in Fig.~\ref{fig:9}), with two antipodal surface defects of topological charge $+1$ in line with the hairy-ball theorem~\cite{Poincaretheorem}. As activity is increased, this \emph{quiescent phase} gives way to another regime, where the droplet spontaneously rotates in steady state (Fig.~\ref{fig:11}(a)). The quiescent droplet has spherical symmetry, whereas it deforms in the rotating phase, attaining the shape of a prolate ellipsoid (Fig.~\ref{fig:11}(a)).
The director field on the droplet surface exhibits bending deformations, typical of extensile suspensions (see Sec.~\ref{sec:3}), that are strongest at the equator. They act as a momentum source (see yellow arrow in Fig.~\ref{fig:11}(a)), hence power stable rotational motion.
In this regime the rotational velocity $\omega$ is linearly proportional to the activity $\zeta$~\cite{Carenza22065}. 
The dynamics is basically driven by the mutual balancing of active injection and viscous dissipation, since elastic absorption is highly suppressed, having the LC attained a stationary pattern. But what if  activity is further increased so that the rate at which energy is injected is not compensated by dissipation?
In this case the excess energy may strengthen elastic deformations, which result into stronger active forces, hence in the increment of the total  kinetic energy.  Active flows are then capable to advect the LC, pulling it apart from its stable orbit, resulting in further deformations leading to further momentum source. The consequent dynamics is no more periodic and the flows become chaotic~\cite{Carenza22065}.
This regime is the droplet analogue of active turbulence~\cite{wensink2012,giomi2015,Doostmohammadi2017,Copar2019,Alert2020,carenza2020_bif}. As such \textcolor{red}{a} chaotic state sets in, the droplets eventualy looses cylindrical symmetry~\cite{Carenza22065} 
and closed disclination lines with hedgehog topological charge $Q=0$ proliferate in the droplet bulk. This behavior, recently reported also in experimental realizations of \emph{unconfined} active nematics~\cite{Duclos2020} is due to the fact that a defect line with null hedgehog charge pays a lower energetic cost than a point defect or a loop with $Q \neq 0$ as we discussed in a previous Section.
These disclination loops undergo a chaotic evolution and eventually merge with other loops evolving in the bulk. 
This phenomenology has important effects on the overall dynamics of the droplet.
Indeed, the asymmetric configuration of LC patterns leads to a net injection of momentum which results into the rambling motion of the droplet~\cite{carenza2020_physA}.

\subsection{Cholesteric droplets: a new route to motility}
\label{sec:chol}
In the previous Sections we have commented on how droplet translational motion occurs as the result of the spontaneous symmetry breaking of the inversion symmetry, while rotational motion arises as the result of spherical symmetry breaking.
Recently, Carenza \emph{et al.} proved that these two dynamical modes can be coupled in a natural way by considering chiral liquid crystals.
Chirality is indeed a generic feature of most biological matter~\cite{ramaswamy2010,Zhou2014,naganathan2014}. A right-left asymmetry may arise at either the microscopic or macroscopic level, and be due to thermodynamic (passive) or non-equilibrium (active) effects. For instance, kinesin or dynein motors exploit ATP hydrolisis to twist their long chains and apply a nonequilibrium active torque on the fibres they walk along~\cite{Wang2012} (see Fig.~\ref{fig:3}(c)). Similarly, bacteria such as~\emph{Escherichia coli}, but also sperm cells, are equipped with long helical flagella. Motor proteins anchored to the cellular membrane generate torques to impart rotational motion on the flagella, whose helix generates a flow in the viscous environment leading to cell propulsion~\cite{Purcell1997,Riedel2005}, as shown by the sketch in Fig.~\ref{fig:3}(d).

Understanding the outcome of the interplay between chirality and activity is therefore an important and timely question and has recently attracted the attention of many scientists in the field~\cite{furthauer2012,furthauer2013,hoffmann2020}. In this Section we will consider a system which is \emph{inherently} chiral and apolar, and so can be modeled --in the passive phase-- as a cholesteric liquid crystal (CLC)~\cite{zumer,FrankPrice}. 
Molecules in the chiral nematic phase twist in the direction perpendicular to the nematic director field. The resulting equilibrium pattern in unconfined geometries is characterized by a helical pattern whose pitch $p_0$ defines the periodicity of the chiral phase and corresponds to the length over which the director field rotates \textcolor{black}{by} $2\pi$ around the helix axis. 
To include these chiral features into the continuous model, the most immediate (and effective) choice is to introduce a suitable term at the level of the elastic free energy to favor twisting deformation:
\begin{equation}
\mathcal{F}^{el}=\dfrac{L}{2}\left[\left( \nabla \cdot \mathbf{Q} \right)^2 + \left[\nabla \times \mathbf{Q} + 2 q_0 \mathbf{Q} \right]^2 \right],
\label{eqn:chol_elastic_energy}
\end{equation}
where the single constant approximation was adopted\footnote{To avoid confusion, it should be clarified that $(\nabla \cdot \mathbf{Q})_\beta = \partial_\alpha Q_{\alpha \beta}$ and $(\nabla \times \mathbf{Q})_{\alpha \beta} =\epsilon_{\alpha \gamma \delta } \partial_\gamma Q_{\delta \beta}$. Moreover, notice that the handedness of the chiral helix is fixed by the sign of $q_0$; in particular a right-handed helix is achieved by choosing $q_0 >0$ while for $q_0<0$ the coiling is left-handed.}. 
Here, $q_0$ is an inverse length proportional to the chirality of the LC and defines the equilibrium pitch $p_0$ of the chiral helix.

In the case of an active cholesteric droplet a particulartly appealing regime was found, fundamentally different from the rotating phase of active nematics. Here, the surface defect pattern is a pair of nearby $+1$ defects (panel (i) of Fig.~\ref{fig:11}), reminiscent of a \emph{Frank-Price structure}~\cite{zumer,FrankPrice,depablo}. 
Interestingly, there is a suggestive analogy between this topological state and a magnetic monopole with its attached \emph{Dirac string}~\cite{kurik1982,depablo} which joins the centre of the droplet with the defect pair.
Such region of maximal layer distortion also represents the region of maximal energy injection where the velocity field is larger (see Fig.~\ref{fig:11}). Under the effect of the activity, the two defects rotate around each other. Remarkably, this time the rotation is accompanied by a translation along the direction of the rotation axis -- thereby resulting in a 
screw-like motion, with the axis of the roto-translation parallel to the \emph{Dirac string}.
This motility mode is compatible with the chiral symmetry of the system, which introduces a generic non-zero coupling between rotations and translations. 

\paragraph{Cholesteric droplets with active torque dipoles: rotation and disclination dance.}

We shall next briefly mention the case of a cholesteric droplet where chirality is added to the model even at a non-equilibrium level due to the action of active torque dipoles. These are able to introduce a nonequilibrium twist in a nematic droplet~\cite{tjhung2011}, whose handedness may reinforce or oppose the handedness of the thermodynamic twist, which is determined by $q_0$. 
This can be done by considering the following active stress term in the Navier-Stokes equation:
\begin{equation}
\sigma_{\alpha \beta}^{act} =  - \bar{\zeta} \epsilon_{\alpha \mu \nu} \partial_\mu (\phi Q_{\nu \beta}).
\end{equation}
This term describes the source of angular momentum provided by torque dipoles that swimmers deploy on the surrounding environment. For positive $\bar{\zeta}$ the active torque corresponds to a pair of out-warding torques (analogue to the one which opens a bottle cap). In turn, $\bar{\zeta}<0$ describes an inward pair of torques (similar to the one used to close a bottle cap). Importantly, the non-equilibrium twist introduced by the activity may reinforce or oppose the handedness of the
thermodynamic twist, which is determined by the sign of $q_0$. Therefore, what is physically meaningful is the relative orientation of the thermodynamic handedness with respect to the one introduced by active torques.
For instance, for the right-handed LC here considered, out-warding active torques ($\bar{\zeta}>0$) result into flows which strengthen the thermodynamic twist. Conversely, a negative active torque would have the opposite effect, countering the handedness of the LC. The opposite would happen in case of a left-handed cholesterics.

In~\cite{Carenza22065} the effect of non-equilibrium chirality on active droplets was considered and different dynamical modes were observed, with the appearence of a number of periodical regimes. 
In particular, for in-warding force dipoles and low chirality the droplet is pierced by two disclination lines which end in +1/2 surface defects. Interestingly, the droplet, fuelled by the active torque, alternates opposite sense of rotation along the same axis in a periodical fashion. Indeed the rotation of the two pairs of surface defects eventually leads to the rewiring of the two disclination lines forming two separate right-handed helices (Fig.~\ref{fig:11}(e)-(h)).

\subsection{Active nematic shells: from morphogenesis to self-assembing materials}
\begin{figure}[th!]
\center
\includegraphics[width=0.8\textwidth]{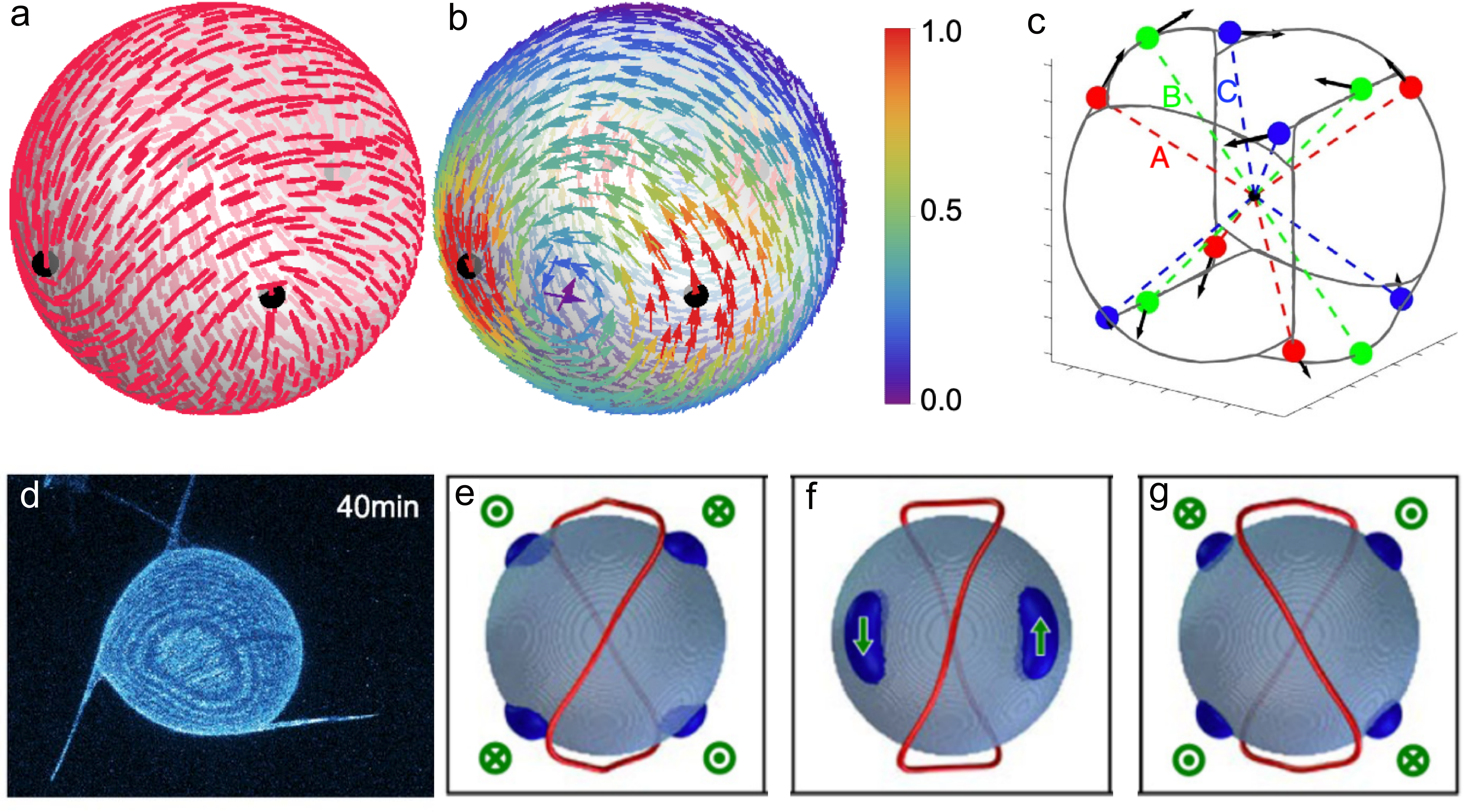}
\caption{ \textbf{Active nematic shells.} \textit{\textbf{(a)-(c)} Shells of active nematics.} Panel (a) shows the orientation field of a tehraedical configuration while panel (b) shows the resulting tangential flow showing the typical vortex structure obtained through the particle description of defects~\cite{Khoromskaia2017}. The four $+1/2$ defects undergo a periodic motion depicted in (c), where defect positions are represented by filled circles  at three consecutive times, marked by A, B and C. The defects orientations are illustrated by arrows, and the dashed lines connect the defects and the origin. \textit{\textbf{(d)} Experimental realization of a an active nematic film on the surface of a spherical vesicle.} Panel (d) shows a confocal image showing the z-projection of the vesicle shape. Four dynamic protrusions grow from the defect sites while the vesicle slowly deswells. \textit{\textbf{(e)-(g)} Active shells emulsified in a passive nematic background.} Green arrows represent the direction of motion of the surface defects (represented in blue) while the red string shows the configuration of the disclination loop of the external passive nematics. The three configurations at different times shows how the periodical motion of the surface defects is able to synchronize the oscillatory dynamics of the Saturn ring. 
\textcolor{black}{Panels (a)-(b) are adapted from~\cite{Khoromskaia2017} with permission from IOP publishing; panel (c) from~\cite{zhang2016} with permission from Springer Nature, panel (d) from~\cite{Keber1135} with permission from L. Giomi, Copyright 2014 American Association for the Advancement of Science, panels (e-g) from~\cite{guillamat2018} with permission from American Association for the Advancement of Science.}}

\label{fig:12}       
\end{figure}

Also active nematic shells have been largely investigated both experimentally and theoretically. Keber \emph{et al.}~\cite{Keber1135} were the first ones to encapsulate microtubule bundles into deformable vescicles to generate a nematic shell. Similarly to the equilibrium case discussed in Section~\ref{sec:noneuclidean} the elastic repulsion between the four $+1/2$ defects (topologically required by the spherical confining geometry) favors a tetrahedral configuration (see Fig.~\ref{fig:9}(e)). However, the defects also propel under the fuelling effect of the activity. The mesmerizing dynamics observed by Keber \emph{et al.} due to the competition between activity and elasticity consists of the periodical motion of the four defects with the tetrahedrical configuration evolving into a planar one and vice versa, as sketched in Fig.~\ref{fig:12}(c). Such oscillatory dynamics can be analytically captured by treating the 4 defects as self-propelled repulsive particles constrained on a sphere whose evolution is governed by an overdamped Newton-Euler equation~\cite{Keber1135,Khoromskaia2017} (see panels (a-b) of Fig.~\ref{fig:12}). 
Keber  \emph{et al.} also observed that the vescicle could be largely deformed and could sustain the formation of long protrusions located in correspondance of the nematic defects (see Fig.~\ref{fig:12}(d)) a phenomenology reminescent somehow of tentacles morphogenesis in the \emph{Hydra}~\cite{Livshits2017,Maroudassacks2021}. These observations suggested that living organism could exploit topological mechanisms to perform their tasks. A review of the literature on this topic goes well beyond the purposes of this Chapter. However, the interested reader can refer to~\cite{Metselaar2019,Hoffmann2021} for recent advances on this subject.
Active nematic shells with the periodic dynamics found by Keber \emph{et al.} were also studied in~\cite{guillamat2018}. Here the active shells were emulsified in a passive nematic background characterized by the formation of a disclination loop (or Saturn ring) as shown in Fig.~\ref{fig:12}(e-g). Interestingly, the periodical motion of the active defects on the shell was also able to synchronize the deformation of the Saturn ring, opening the way towards new possibilities in the field of self-assembling materials and development of biologically based micro-motors. 

To conclude we report that recently the case of deformable active shells and membranes was taken into account in~\cite{Metselaar2019,Hoffmann2021,Vafa2021}. In particular, in~\cite{Metselaar2019} the relation between active topological defects and the local curvature of the shell was considered and suggested physical insights regarding how morphogenesis in living organisms might be actually driven by topology.

\section{Aknowledgement}
The authors thank Luca Giomi, Davide Marenduzzo and Enzo Orlandini for many useful discussions.



\bibliography{biblio.bib} 
\bibliographystyle{rsc}

\end{document}